\documentclass[%
 aip,
 amsmath,amssymb,
reprint,%
]{revtex4-1}

\usepackage{graphicx}
\usepackage{dcolumn}
\usepackage{bm}

\usepackage[utf8]{inputenc}
\usepackage[T1]{fontenc}
\usepackage{mathptmx}
\usepackage{etoolbox}
\usepackage[version=4]{mhchem}
\usepackage{array}
\usepackage{tabularx}
\makeatletter
\def\@email#1#2{%
 \endgroup
 \patchcmd{\titleblock@produce}
  {\frontmatter@RRAPformat}
  {\frontmatter@RRAPformat{\produce@RRAP{*#1\href{mailto:#2}{#2}}}\frontmatter@RRAPformat}
  {}{}
}%
\makeatother

\preprint{AIP/123-QED}

\date{\today}

\usepackage{ulem}
\usepackage[dvipsnames]{xcolor}
\usepackage[colorlinks=true,citecolor=Cerulean,linkcolor=RubineRed,urlcolor=Cerulean]{hyperref}
\usepackage{amsmath,amssymb,amsbsy,amsfonts,amsthm}
\usepackage{fancyhdr}
\usepackage{graphicx,subfigure}
\usepackage{epsfig}
\usepackage[svgnames]{xcolor}
\usepackage{bbm}

\usepackage{multirow}
\usepackage{longtable}
\usepackage{booktabs} 
\usepackage{orcidlink}  

\begin{document}

\title{Dynamic hysteresis in an autocatalytic reaction network}

\author{Sheela Yadav}
\affiliation{Indian Institute of Technology Mandi, Kamand, Himachal Pradesh 175075, India
}

\author{Jason~R.~Green\orcidlink{0000-0003-2572-2838}}
\email[]{jason.green@umb.edu}
\affiliation{Department of Chemistry,\
University of Massachusetts Boston,\
Boston, MA 02125
}
\affiliation{Department of Physics,\
	University of Massachusetts Boston,\
	Boston, MA 02125
}

\author{Moupriya Das\orcidlink{0000-0003-1851-2162}}
\email[]{moupriya@iitmandi.ac.in}
\affiliation{Indian Institute of Technology Mandi, Kamand, Himachal Pradesh 175075, India
}


\begin{abstract}

A periodically driven chemical network can be expected to exhibit memory of past driving conditions. 
Here we show that an autocatalytic reaction network can exhibit dynamic hysteresis 
as a result of the competition between its intrinsic relaxation time scale and that of the periodic drive.  
The autocatalytic reaction steps generate bistability in the concentration of the autocatalytic species, and periodic pumping of the product species provides the external drive. 
Hysteresis arises from the lag between the concentration response of the autocatalytic species and the external periodic drive. 
The resulting hysteresis-loop area quantifies the extent of the system's hysteretic response.
Using the periodically driven Schl\"ogl model as a representative bistable chemical system, we use this loop area to determine how the magnitude of the hysteretic response is controlled by the driving protocol, intrinsic fluctuations, and the size of the system. 
Varying the driving frequency and the strength of the fluctuations causes a turnover of the hysteresis loop area, whereas the area changes monotonically with driving amplitude and system size. 
These trends identify the driving protocol and fluctuation strength as primary controls on the magnitude of dynamic hysteresis. 
Unlike static hysteresis, the dynamic hysteresis loop area tends to vanish in the quasistatic limit of very slow periodic driving. 
In contrast to stochastic resonance, which is typically associated with weak periodic forcing and noise-assisted amplification, dynamic hysteresis can characterize the reaction-system response over a wider range of external control conditions. 
We further show that the delayed concentration response is mirrored in Shannon entropy and in the total entropy production rate, connecting dynamic hysteresis to information-theoretic and stochastic-thermodynamic measures of irreversibility. Overall, we identify and interpret the role of the controlling factors in chemical dynamic hysteresis and suggest implications for efficient chemical logic gates and eventually chemical computers. 
\end{abstract}

\maketitle

\section{Introduction}
\label{sec: intro}

Chemical changes occur away from equilibrium and are subject to fluctuations~\cite{Remlein2024}. 
When driven periodically~\cite{leonard1994stochastic}, chemical reaction networks may retain memory of previous driving conditions through delayed changes in species concentrations~\cite{wong2022}. These networks exhibit a wide variety of phenomena under nonequilibrium driving, including multistability, oscillatory dynamics, and enhanced entropy production, demonstrating how nonequilibrium driving fundamentally affects complex dynamics~\cite{nicolaou2023prevalence}. 
In autocatalytic networks~\cite{delbruck1940,nicolis1977,Vanag1999}, nonlinear feedback can support distinct low- and high-concentration states of the autocatalyst, and a time-periodic drive can push the system between these states~\cite{leonard1994stochastic}.
Because transitions between the two concentration regimes require a finite relaxation time, it is possible for the system's response to lag behind the external drive.
This phase lag produces a history-dependent relation between the drive and the concentration response, which appears as dynamic hysteresis; the magnitude of this response is determined by the competition between external modulation and intrinsic relaxation~\cite{mahato1994hysteresis,Das2012_MD}. 

Hysteresis is a phenomenon in which the system's response depends on the history of the controlling parameter. It has been observed in physical~\cite{mahato1994hysteresis,chakrabarti1999dynamic,casteels2016dynamic,Dykman1994PRL,Gammaitoni1998,Ghosh2025,Paul2021}, chemical~\cite{ross1976,Pal2025,guidi1997bistability,Guidi1998,Schiffmann1982,Ball2001} 
biological~\cite{pajaro2019transient,Kim2012,Aguda2003, jiang2019single}, and mechanical systems~\cite{wang2023model,lynch2025hysteresis}.
Dynamic hysteresis is a nonequilibrium response in which a system, periodically driven in time by an external energy source, follows different paths during the forward and reverse portions of a driving cycle~\cite{Tierno2013,das2012dynamical}.
Unlike static hysteresis, the loop area is dynamic hysteresis tends to vanish in the quasistatic limit of very slow modulation~\cite{chakrabarti1999dynamic,Banerjee2015}. 
Understanding how to control dynamic hysteresis is important as it occurs in systems ranging from magnetic~\cite{sun2026universal,Rao1990}, electrical~\cite{Jo2007, Tagantsev2001}, electronic~\cite{Bisquert2023, Bisquert2024PRX}, to mechanical devices~\cite{AlBender2004, Danilin2017,Laurson2012} 
Dynamic hysteresis in chemical systems has not been thoroughly analyzed. Some
experimental examples include thermal and chemical hysteresis~\cite{yamaguchi2021thermal,dadi2016, raj2015steady}, adsorption-desorption isotherms~\cite{Edison2021,sarkisov2000hysteresis,monson2012}, and ion transport or interfacial processes in photovoltaic devices~\cite{contreras2016specific,tress2015,snaith2014,richardson2016}. 
A systematic theoretical understanding is still needed to control and design dynamic hysteresis in chemical systems, especially for stochastic chemical reaction networks.

The Schl\"ogl model~\cite{Schlogl1972} provides a minimal model for this problem because it combines autocatalysis, intrinsic bistability~\cite{prakash1997}, and stochastic transitions between metastable concentration states~\cite{Wang2017,verma2018stochastic}. 
This model was particularly useful in developing our understanding of stochastic resonance. 
While stochastic resonance focuses on how weak periodic forcing and intrinsic fluctuations cooperate to enhance transitions between metastable concentration states~\cite{leonard1994stochastic}, dynamic hysteresis requires characterizing the delayed response over a larger range of driving amplitudes and frequencies~\cite{QuondamAntonio2022}. 
What remains unclear is how the magnitude of dynamic hysteresis in a driven stochastic chemical network is controlled by driving frequency, driving amplitude, intrinsic fluctuations, and system size. 

Here we examine dynamic hysteresis in the periodically driven Schl\"ogl model, a minimal autocatalytic reaction network with intrinsic bistability. 
The external control is implemented through periodic pumping of a product species, and the response is the concentration of the autocatalytic species.
We use numerical simulations of the chemical Langevin dynamics~\cite{Kim2017,Ma2016}, which include intrinsic fluctuations in concentration. We quantify the response through the hysteresis-loop area as a function of driving frequency, driving amplitude, system size, and well width (a measure of the fluctuation strength). 

Chemical logic operations, and consequently chemical computational frameworks, can be built with autocatalytic reaction networks~\cite{Egbert2019,Arcadia2021,Peng2024,kriukov2024}, suggesting the results here might contribute to the understanding of the logical response of chemical systems to logical inputs. In practice, logically irreversible operations can also be physically irreversible.
In driven systems, hysteresis loss is also often associated with dissipation and entropy production~\cite{Tshiprut2009,Santamaria2014,Sasso2006,Li2017ECM}.
For the present chemical network, this motivates a second issue: how information-theoretic and stochastic-thermodynamic quantities quantify the delayed concentration response.
We therefore compare hysteresis in the ensemble-averaged autocatalyst concentration with hysteresis in Shannon entropy and in the total entropy production rate.
The resulting loop areas show similar dependence on control parameters, including the driving frequency and the fluctuation strength. 

The remainder of the paper is organized as follows:
In Sec.~\ref{sec: system}, we introduce the Schl\"ogl reaction network and its Langevin description with external periodic driving and state-dependent multiplicative noise.
In Sec.~\ref{sec: results}, we analyze the concentration response and quantify the dynamic hysteresis loop area under variations in driving frequency, driving amplitude, system size, and well width.
In Sec.~\ref{sec: stochastic thermodynamic analysis}, we examine the corresponding hysteresis in Shannon entropy and in the total entropy production rate. We distill these results into our main conclusions in Sec.~\ref{sec: conclusion}.

\section{System and dynamics}
\label{sec: system}

To analyze dynamic hysteresis in a minimal autocatalytic network, we consider the reaction scheme of the Schl\"ogl model~\cite{leonard1994stochastic}:
\begin{subequations}\label{eq:reactions}
\begin{align}
\ce{A} + 2\ce{X} &\underset{k_2}{\stackrel{k_1}{\rightleftharpoons}} 3\ce{X} \label{eq:reactions_a} \\
\ce{B} + \ce{X} &\underset{k_4}{\stackrel{k_3}{\rightleftharpoons}} \ce{C}  \label{eq:reactions_b}
\end{align}
\end{subequations}
We treat $\ce{A}$ and $\ce{B}$ as chemostatted reservoirs with fixed concentrations $A$ and $B$, and impose the periodic drive through the externally controlled concentration $C$ of $\ce{C}$; thus, only the population $n$ of the autocatalytic species $\ce{X}$ evolves stochastically. 
Because molecular populations fluctuate in finite chemical systems, the dynamics of species $\ce{X}$ are naturally described first at the level of a chemical master equation.

The probability $P_n(t)$ of having $n$ molecules of species X at time $t$ evolves according to the one-step chemical master equation for this birth-death process~\cite{gillespie1977exact,vanKampen2007}: 
\begin{equation}\label{eq:master_eq}
\frac{dP_{n}(t)}{dt} = t_{n-1}^+(t)P_{n-1}(t) + t_{n+1}^-(t)P_{n+1}(t) - [t_{n}^+(t) + t_{n}^-(t)]P_{n}(t).
\end{equation}
The transition rates, or propensities, $t_n^+(t)$ and $t_n^-(t)$ correspond to birth events $n\rightarrow n+1$ and death events $n\rightarrow n-1$, respectively; the polynomial factors ensure that reactions requiring multiple $\ce{X}$ molecules have zero propensity when too few are present: 
\begin{subequations}\label{eq:transition_prob}
\begin{align}
t_{n}^+(t) = \alpha n(n - 1) + \gamma(t), \label{eq:transition_prob_a}\\
t_{n}^- = n(n-1)(n-2) + \beta n. \label{eq:transition_prob_b}  
\end{align}   
\end{subequations}
The propensities in Eq.~(\ref{eq:transition_prob}) define the reduced parameters and the external driving protocol. The dimensionless parameters are $\alpha = k_1 A V/k_2$ and $\beta = k_3 B V^2/k_2$. Here, $A$ and $B$ are the fixed concentrations of the chemostatted species $\ce{A}$ and $\ce{B}$, respectively. The parameter $\gamma$ represents the externally controlled pumping of species $\ce{C}$ through its concentration $C$ and is expressed as $\gamma = k_4 C V^3/k_2$. To incorporate the effect of external periodic driving, we take $\gamma(t)=\gamma_0+\epsilon_0 \cos(2\pi f t)$. Here $\gamma_0$ denotes the constant component of the driving force, related to a constant concentration $C_{0}$ of $\ce{C}$. $\epsilon_{0}$ is the amplitude of the periodic modulation of $C$, and $f$ represents the driving frequency. 

We nondimensionalize time using $\tau=t k_2/V^2$, where $t$ is physical time and $V$ is the volume of the system representing its size. In the following, $t$ denotes this dimensionless time unless otherwise stated. This volume-dependent scaling is consistent with mass-action kinetics, for which stochastic transition rates scale as $1/V^{m-1}$ for a reaction of molecularity $m$~\cite{vellela2009stochastic}. Under this rescaling, time, concentrations, and rate constants are absorbed into the reduced variables and parameters of the model.

A Kramers-Moyal expansion of the chemical master equation yields a Fokker-Planck approximation, which has a corresponding Langevin representation for the concentration variable $x=n/V$. In this one-dimensional description, the stochastic dynamics reduce to overdamped motion in an effective quartic double-well landscape~\cite{leonard1994stochastic}. The Langevin equation is
\begin{equation} \label{eq: stochastic_diff_eq}   
\frac{dx}{dt} = F(x, t)+\frac{1}{V}g(x, t)\xi(t).
\end{equation}
The above Langevin dynamics follow the It\^o convention, as its derivation initiates from the Kramers-Moyal expansion of the chemical master equation \cite{Gillespie2000}. 
Here, the discrete molecular population is represented by the continuous concentration $x(t)$, and intrinsic fluctuations enter through the state-dependent noise amplitude $g(x,t)$. The stochastic term $\xi(t)$ represents intrinsic chemical fluctuations with $\langle \xi(t) \rangle=0$ and $\langle \xi(t)\xi(t') \rangle=2\delta(t-t')$. 
The product $g(x,t)\xi(t)$ is therefore multiplicative noise because its amplitude depends on the state variable $x$~\cite{Doering2005, Phillips2025}; it is also explicitly time-dependent through the periodic driving term.
The deterministic drift is
\begin{equation} \label{eq: deterministic}
F(x, t) = -V^2x^3 + (\alpha + 3)Vx^2 - (\alpha + \beta + 2)x + \frac{1}{V}\gamma(t).
\end{equation}
The noise amplitude is
\begin{equation} \label{eq: noise}
g(x, t) = \sqrt{V^3x^3+(\alpha + 3)V^2x^2+(\alpha+\beta+2)Vx+\gamma(t)}.
\end{equation}
To obtain a symmetric bistable form for the deterministic dynamics, we choose:
\begin{eqnarray} \label{eq: forms}
\alpha &=& 3(Vx_{0} -1), \label{eq: forms_a} \nonumber\\ 
\beta &=& 3(V^2x^2_{0} - Vx_{0} + 1) - w^2 - 2,  \nonumber \label{eq: forms_b}\\
\gamma(t) &=& V^3x^3_{0} - w^2Vx_{0} + \epsilon_{0} \cos(2\pi ft). \label{eq: forms_c}   
\end{eqnarray}
Substitution into Eqs.~(\ref{eq: deterministic}) and (\ref{eq: noise}) gives:
\begin{align}
    F(x,t) = -V^2 (x - x_0)^3 + w^2 (x - x_0) + \frac{1}{V}\epsilon_{0} \cos(2\pi f t).
    \label{eq: force_forms}
\end{align}
\begin{align}
    g(x,t) = \sqrt{\,V^3 (x + x_0)^3 - w^2 V(x + x_0) + \epsilon_{0} \cos(2\pi f t)}.
    \label{eq: noise_forms}
\end{align}
The parameters in Eqs.~(\ref{eq: force_forms}) and (\ref{eq: noise_forms}) determine the bistable landscape and fluctuation strength. Without the periodic control in the drift term, the potential corresponding to the deterministic drive, $U(x)=\frac{V^2}{4}(x-x_0)^4-\frac{w^2}{2}(x-x_0)^2$, establishes the form of the potential landscape as a symmetric double-well. The system size $V$ controls the strength of chemical fluctuations, $w$ governs the quantitative structure of the bistable potential by influencing both the well separation $2w/V$ and the barrier height $w^4/(4V^2)$, and $x_{0}$ denotes the barrier position. 

The concentration $x$ of the autocatalytic species represents the state of the system. Therefore, each $x$ value on the potential corresponds to the possible concentration of species $\ce{X}$. The two minima at $x_{0} \pm\frac{w}{V}$ represent the stable states or the steady states. They are separated by a metastable state at $x_{0}$ corresponding to the maximum of the potential. The states of the system at the two wells of the double-well potential signify the low and high concentration regimes of the autocatalytic species. The smaller concentration domain of the autocatalyst is the left well, and its larger range is the right well of the potential. The maximum of the potential partitions the low- and high-concentration states of the autocatalytic species. The time-dependent drive $\epsilon_{0}\textrm{cos}(2 \pi f t)$ in the drift term tilts the positions of the minima of the double-well potential periodically in time. 

For the representative parameter values $V=1.01$, $w=0.9$, $x_0=1$, and $\epsilon_0=0.08$, the rates $\alpha$, $\beta$, and $\gamma$ are positive. For the corresponding time-independent deterministic double-well landscape, the barrier height is $0.1608$, and the left and right minima are located at $0.1089$ and $1.89$, respectively. The noise amplitude $g(x,t)$ is state dependent. For the chosen parameters, it is smaller in the left well, corresponding to the low-concentration region of $\ce{X}$, and larger in the right well, corresponding to the high-concentration region.
Together, the deterministic drift, periodic driving, and intrinsic fluctuations generate transitions between the two wells and produce an oscillatory response in $x(t)$ that creates the potential for dynamic hysteresis.

\begin{figure}
      \centering
      \includegraphics[width=0.6\linewidth]{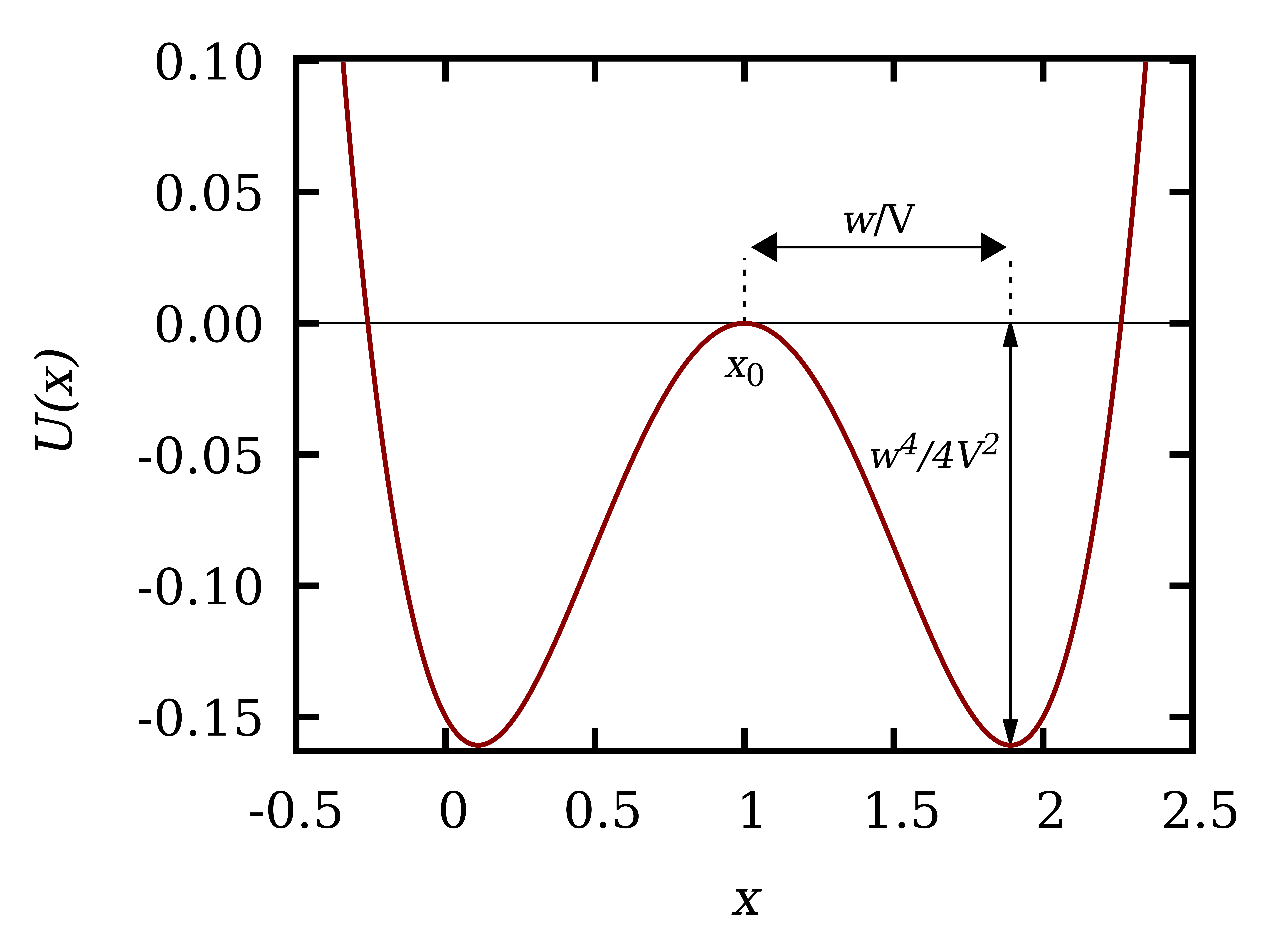}
	  \caption{Bistable potential $U(x)=\frac{V^2}{4}(x-x_0)^4-\frac{w^2}{2}(x-x_0)^2$ with parameter values $V = 1.01$, $w = 0.9$, and $x_{0} = 1$.}  
      \label{fig:bistable}
\end{figure}

In the numerical simulations, we analyze the response $\langle x(t)\rangle$ under the periodic modulation $\epsilon_0\cos(2\pi f t)$ while varying the main control and system parameters: the driving frequency $f$, driving amplitude $\epsilon_0$, potential-well width $w$, and system size $V$. The Langevin equation, Eq.~(\ref{eq: stochastic_diff_eq}), is solved using Heun's method, with Gaussian noise generated by the Box-Muller algorithm. Unless otherwise stated, simulations use a time step of $10^{-4}$ and averages over $100$ cycles and $2\times 10^6$ trajectories. 

\section{Results and Discussion}
\label{sec: results}

We first characterize dynamic hysteresis through the response of the autocatalyst concentration to the periodic drive. In the driven Schl\"ogl model, periodic pumping of the product species controls the concentration $x(t)$ of the autocatalytic species. In the absence of driving, the system has low- and high-concentration steady states; under periodic driving, the autocatalyst concentration is driven between these regimes. Because intrinsic fluctuations affect individual trajectories, we use the ensemble average $\langle x(t)\rangle$ as the primary response observable.

\begin{figure}
        \centering
      \includegraphics[width=0.48\linewidth]{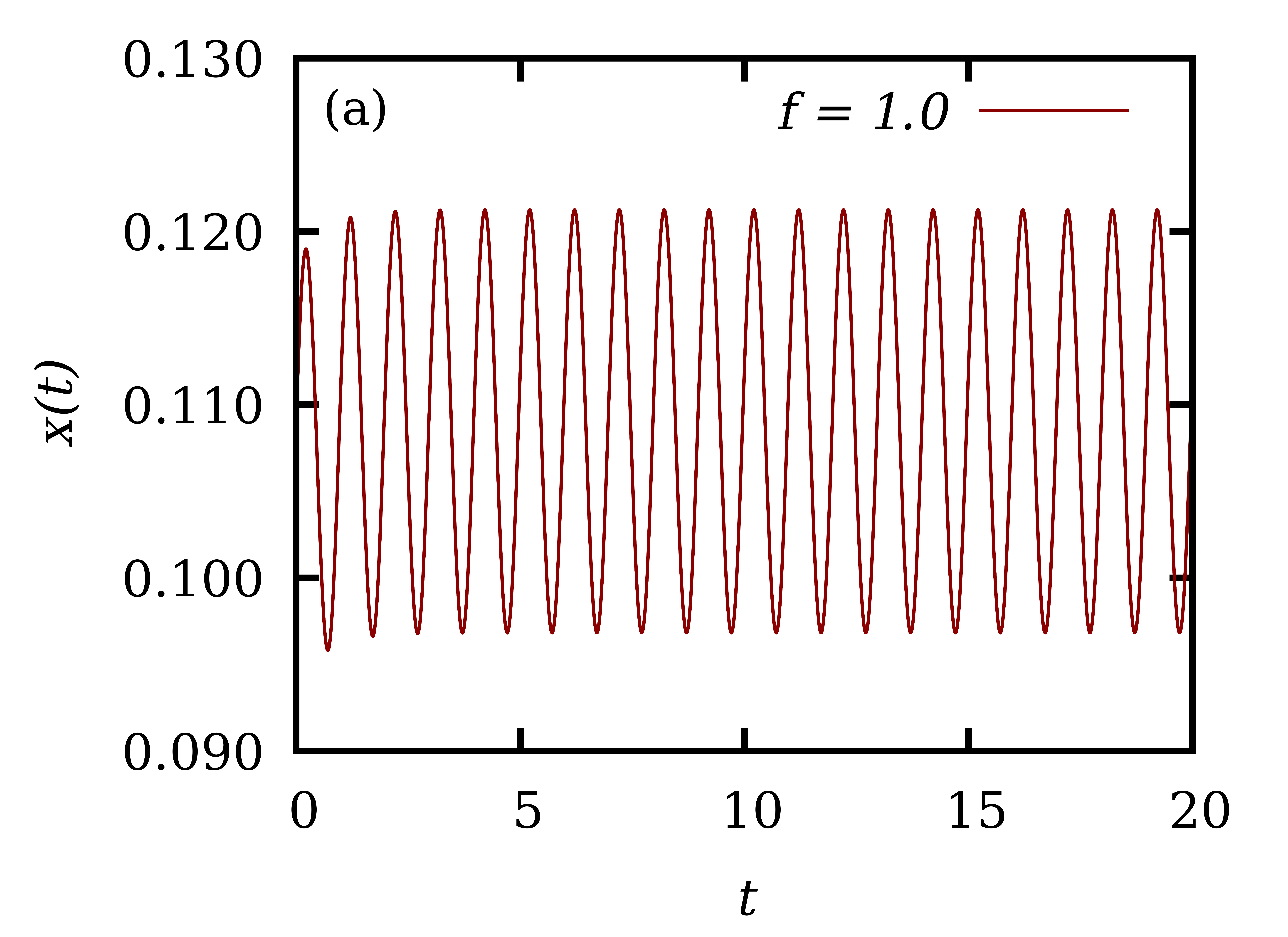}
      \includegraphics[width=0.48\linewidth]{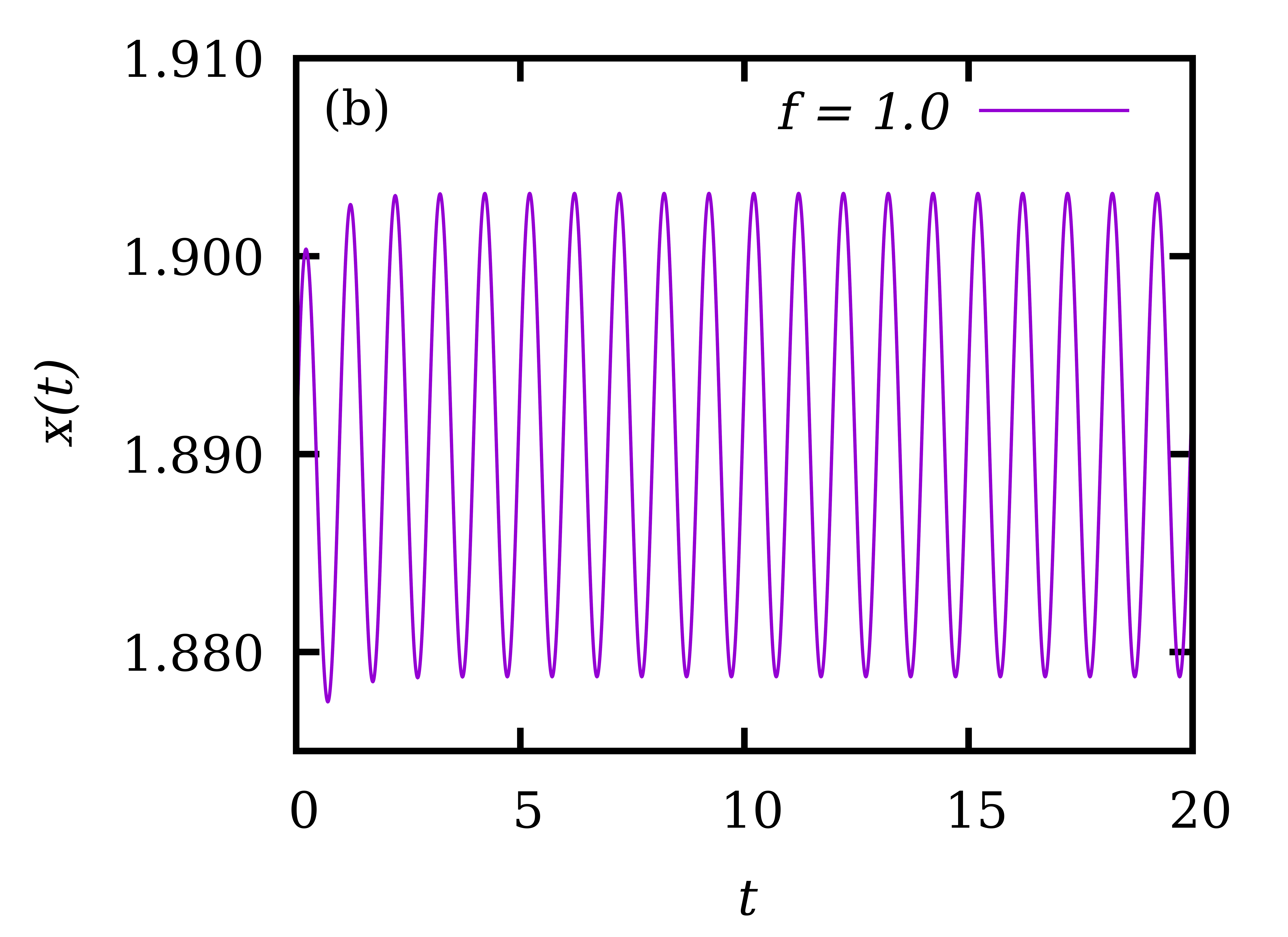}
      \includegraphics[width=0.48\linewidth]{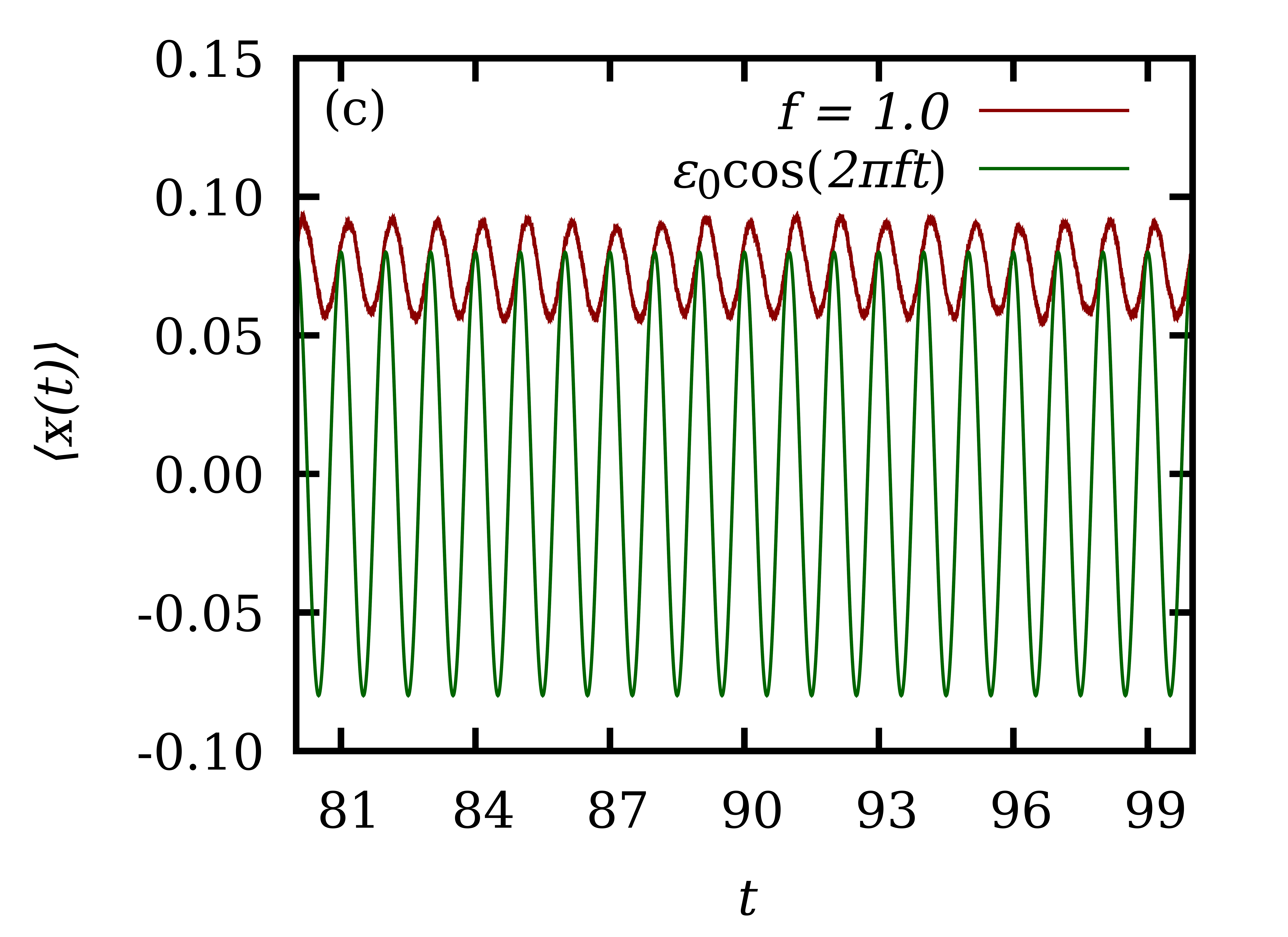}
      \includegraphics[width=0.48\linewidth]{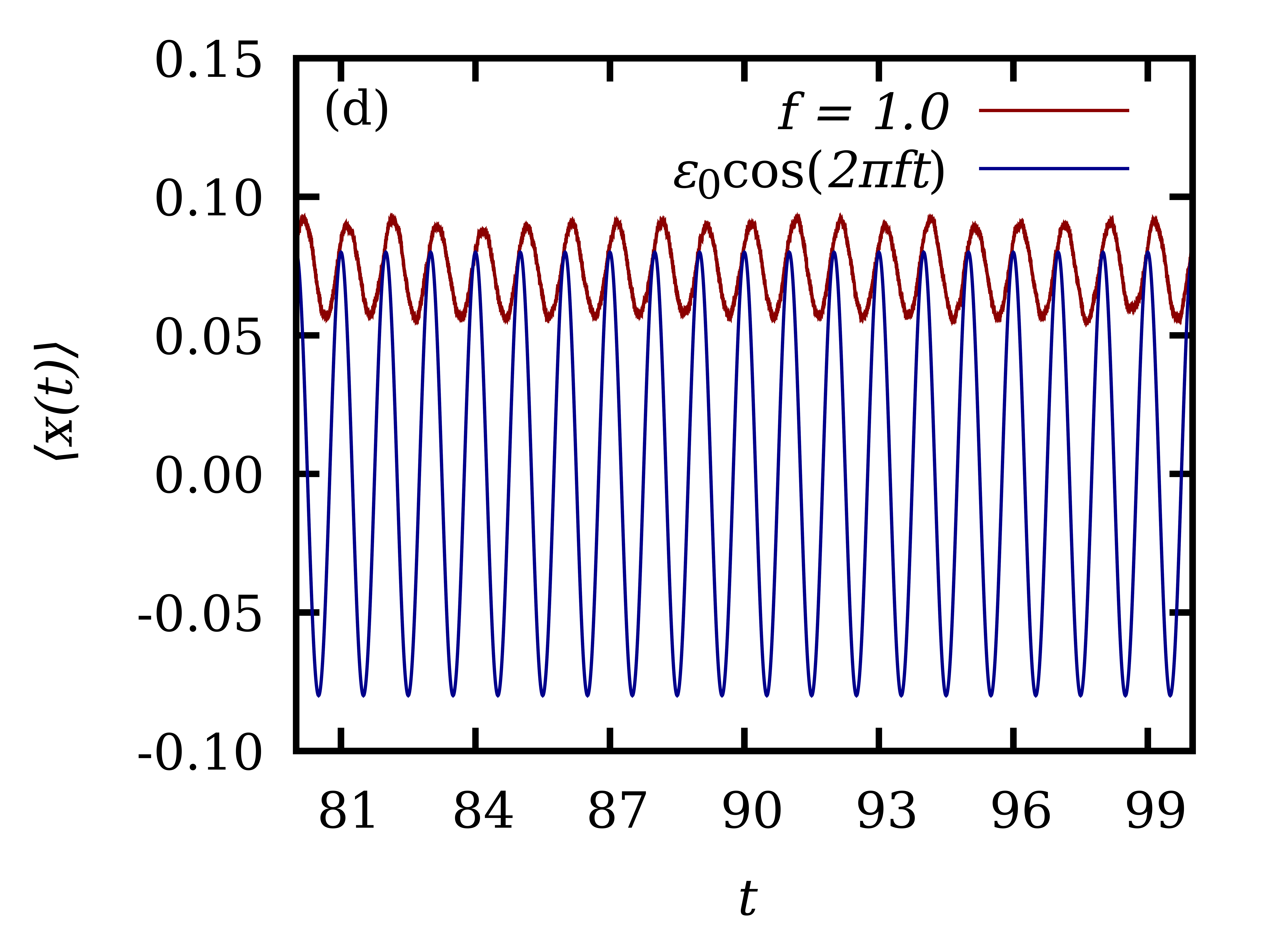}
        \caption{(a-b) Temporal evolution of $x(t)$ initially positioned at the left and right well minima, respectively, under the deterministic drift force in the absence of noise. (c) The time series of ensemble-averaged response $\langle x(t) \rangle$ (red line) of the stochastic system subject to drift, fluctuations, and periodic drive, along with the comparison with the periodic drive (green smooth line) at frequency $f=1$, with the initial condition at the left well minimum. (d) Same as (c) but with the initial condition in the right well; the periodic forcing is shown as the blue smooth curve.}
        \label{fig: time_series}
\end{figure}

For comparison with the stochastic response, we also simulate deterministic trajectories. For the parameter set defined above, trajectories are initialized at one of the well minima unless otherwise stated. Fig.~\ref{fig: time_series}(a) and (b) show representative trajectories governed only by the deterministic drift, without noise. In this limit, the system approaches one of the steady states, depending on the initial condition, and the periodic drive produces oscillations around that state.

\begin{figure}
        \centering
      \includegraphics[width=0.48\linewidth]{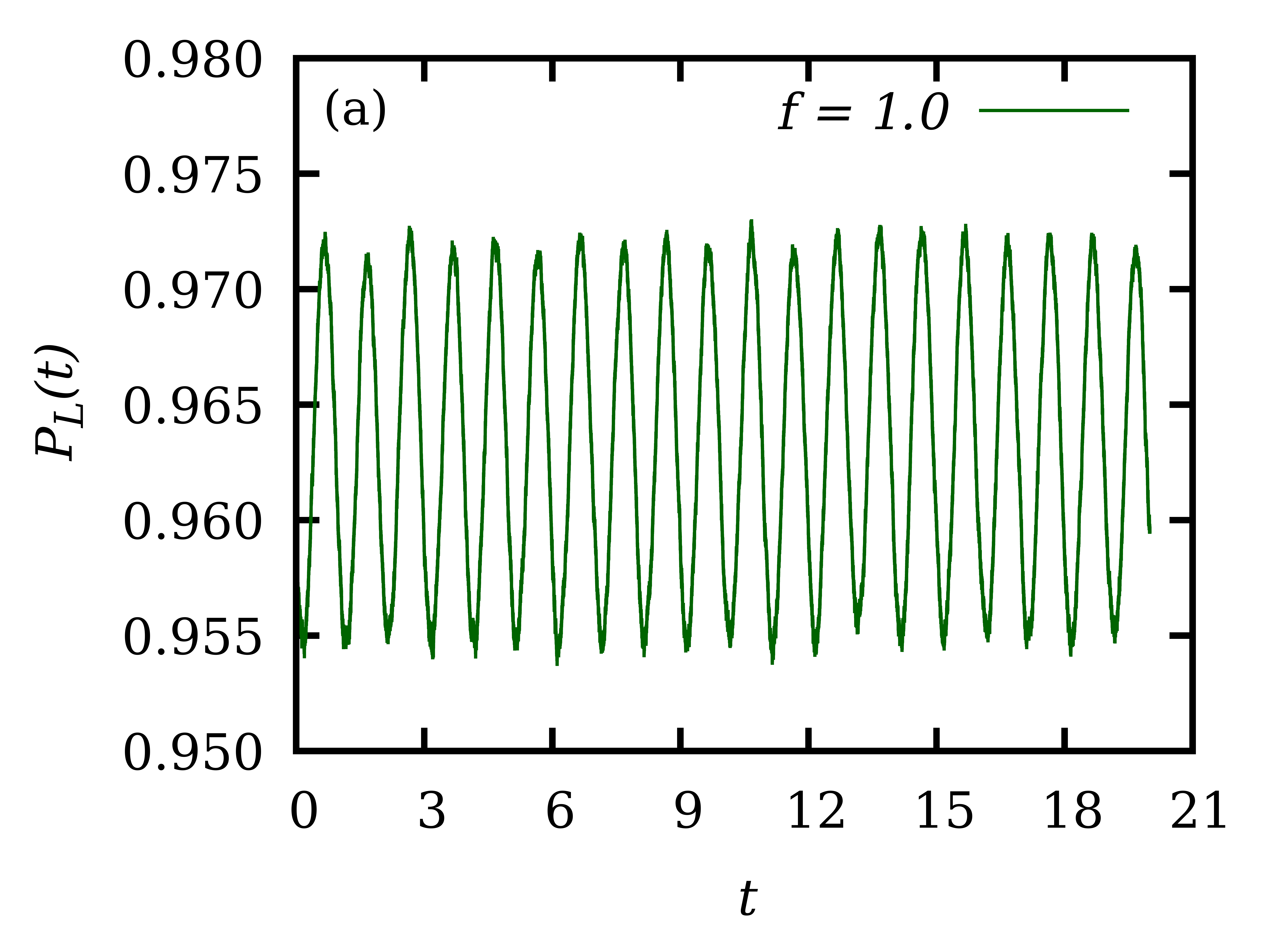}
      \includegraphics[width=0.48\linewidth]{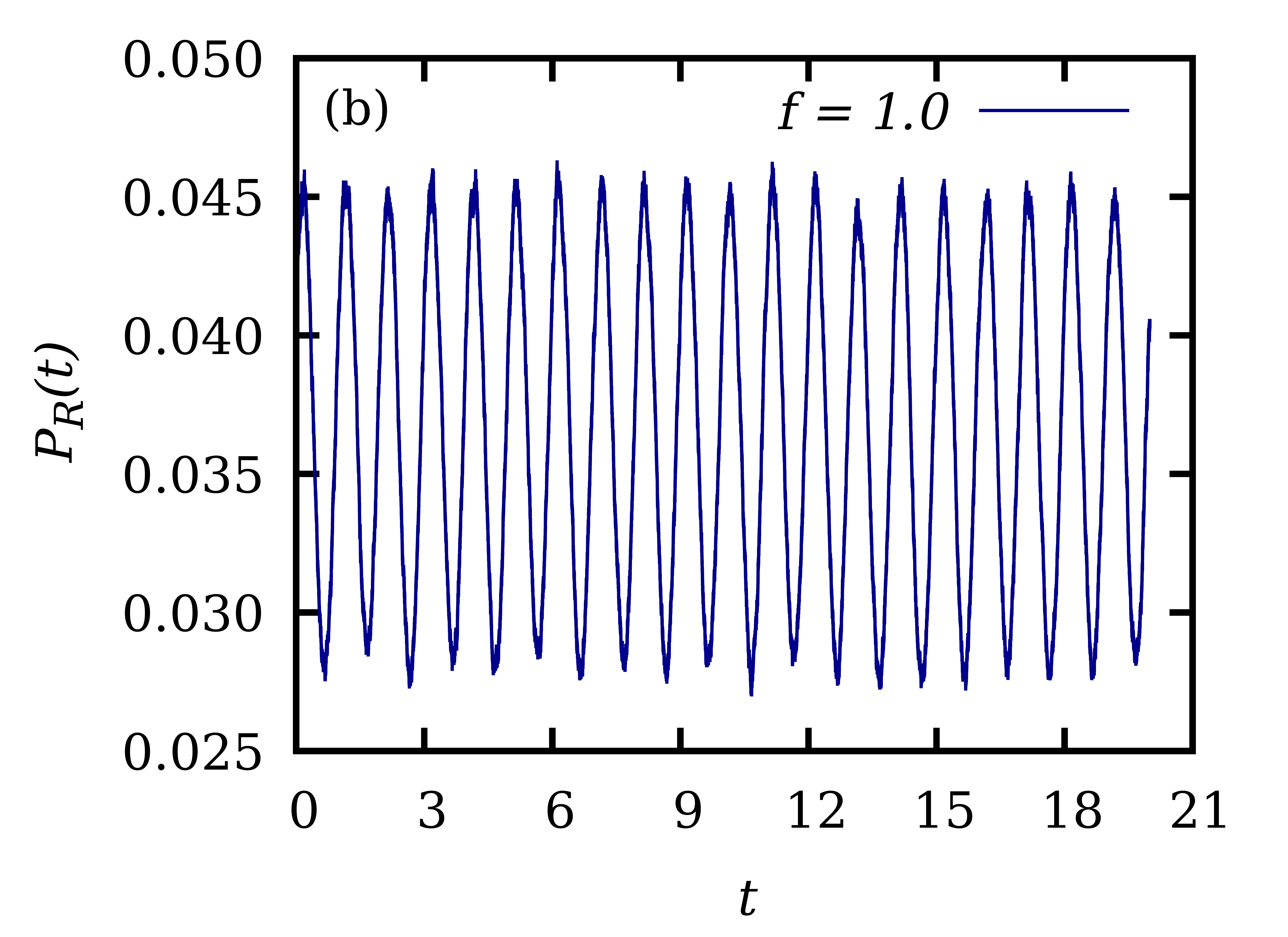}
        \caption{(a-b) Temporal variation of the occupation probabilities for the left ($P_L(t)$) and right ($P_R(t)$) wells at frequency $f = 1$.}
        \label{fig: prob_left_right}
\end{figure}

To measure the stochastic response, we ensemble-averaged concentration over intrinsic fluctuations. Fig.~\ref{fig: time_series}(c) and (d) show $\langle x(t)\rangle$ together with the applied periodic control, for trajectories initialized in the left and right wells, respectively. In both cases, the ensemble response remains mostly in the low-concentration regime. This bias reflects the state-dependent noise amplitude: fluctuations are stronger in the right well and weaker in the left well, making the low-concentration regime more stable on average. The integrated probabilities in the left and right wells, shown in Fig.~\ref{fig: prob_left_right}(a) and (b), support this interpretation.
\subsection{Dynamic hysteresis}
\label{subsec: hysteresis}
A dynamic hysteresis loop provides a direct measure of the delayed nonequilibrium response of the system to periodic driving. Such hysteretic behavior has been investigated in driven colloidal systems,~\cite{Tierno2013,mcdermott2013dynamic,RodriguezGallo2021} while the dependence of hysteresis-loop shape and area on the driving conditions has been extensively studied in kinetic Ising models\cite{sides1998stochastic,rikvold2000dynamic}.  
Our aim is to understand this control of the driving protocol and the system parameters on dynamic hysteresis in a chemical reaction network, using a driven Schl\"ogl reaction as a paradigmatic example. 

\begin{figure*}[ht]

\centering
    
    \includegraphics[width=0.49\linewidth, height=5.5cm]{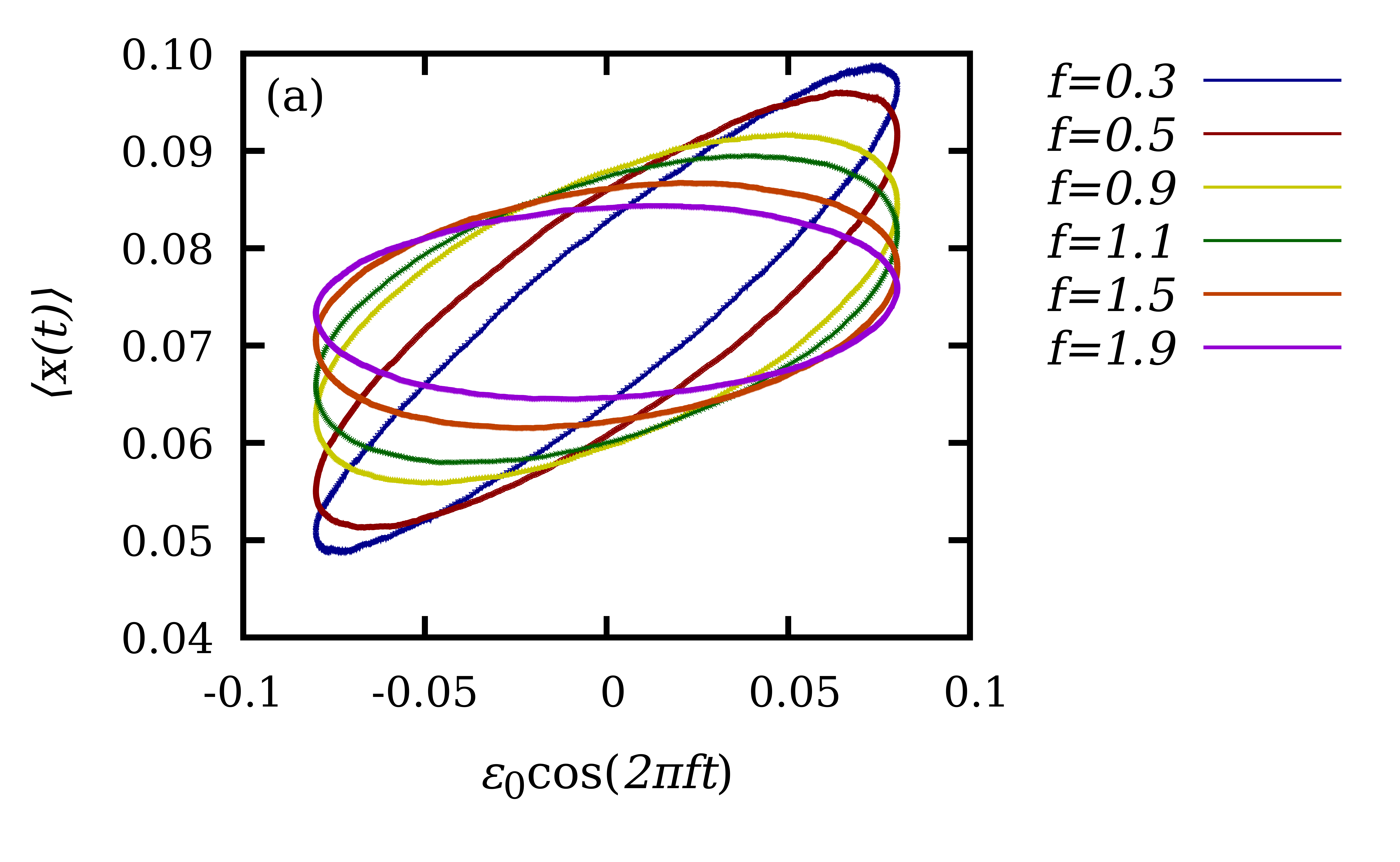}
    \includegraphics[width=0.49\linewidth, height=5.5cm]{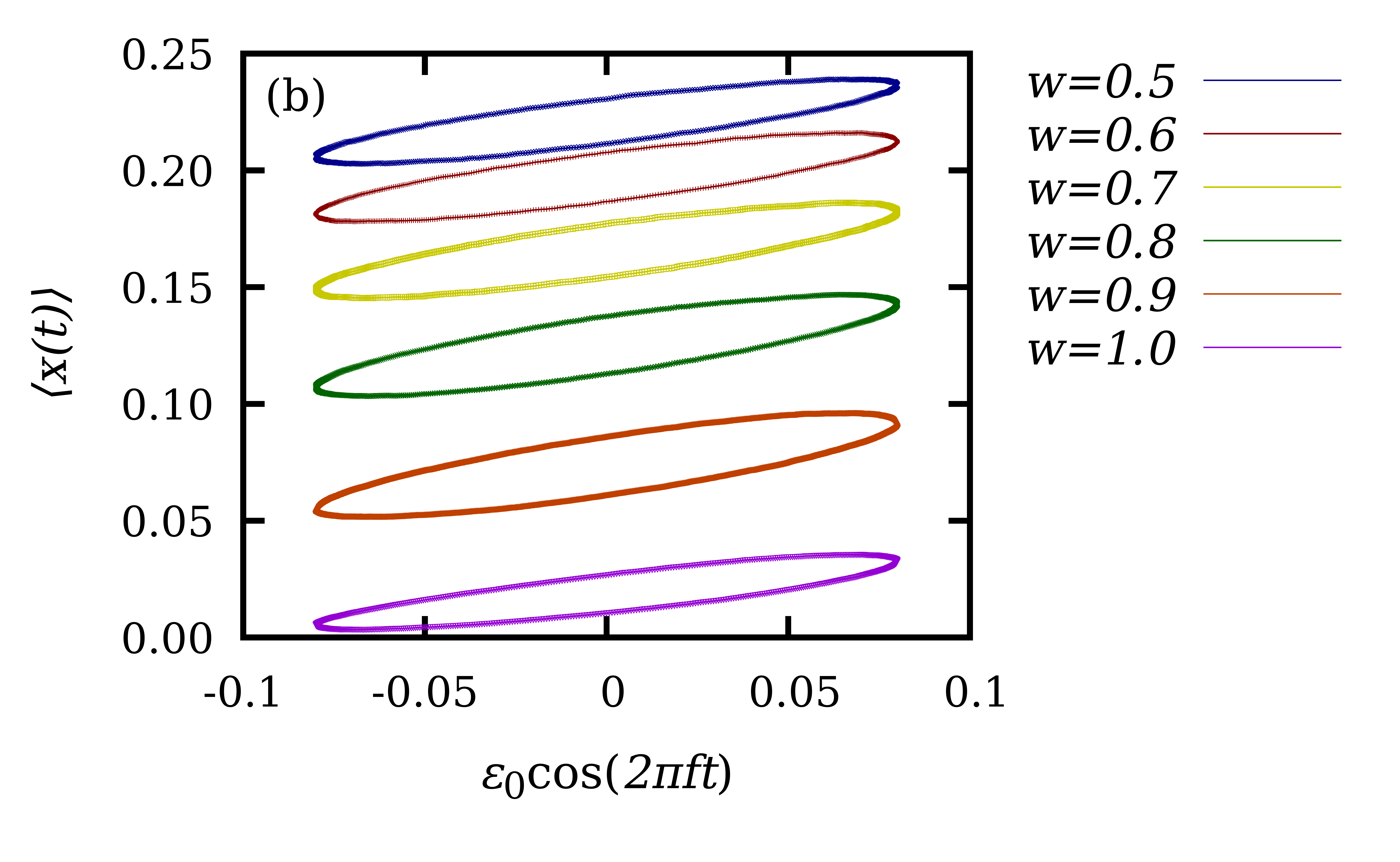}

\vspace{0.25cm}
  
    \centering

    \includegraphics[width=0.49\linewidth, height=5.5cm]{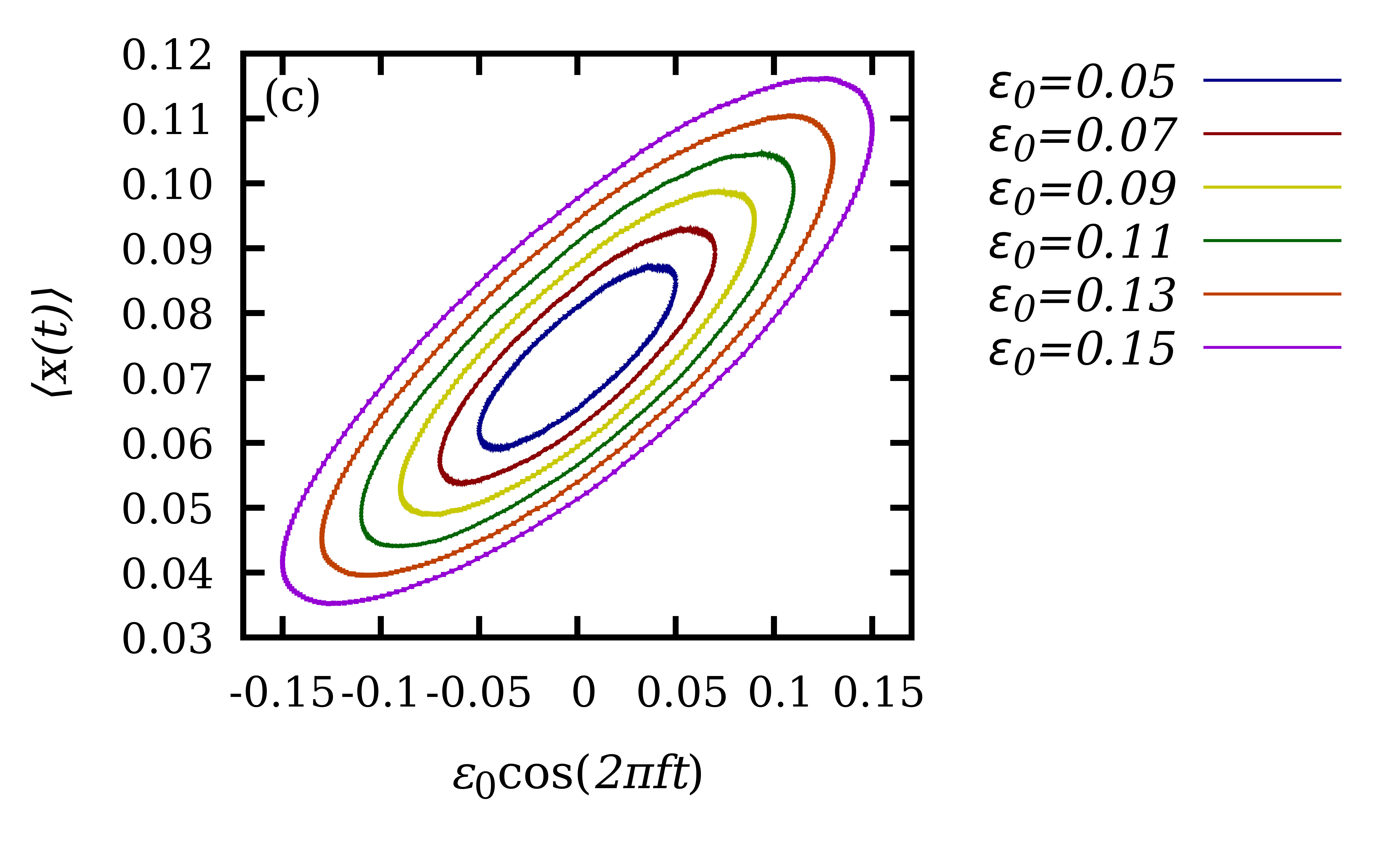}
    \includegraphics[width=0.49\linewidth, height=5.5cm]{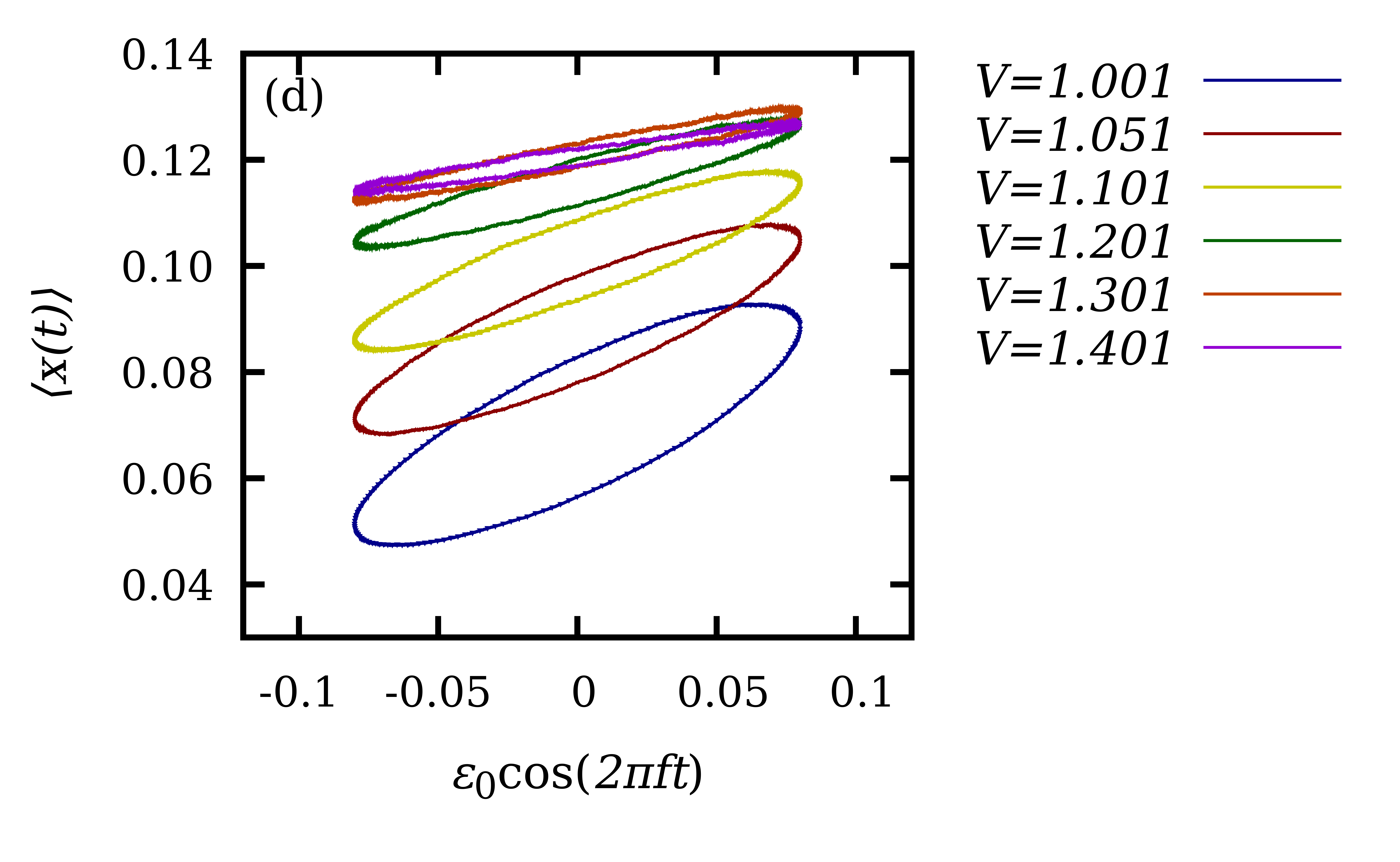}
    \caption{Dynamic hysteresis in the ensemble-averaged autocatalyst concentration. (a) Hysteresis loops of $\langle x(t)\rangle$ versus the periodic drive $\epsilon_0\cos(2\pi f t)$ for different driving frequencies $f$. (b) Hysteresis loops for different well widths $w$ at fixed $f=0.5$. (c) Hysteresis loops for different driving amplitudes $\epsilon_0$ at fixed $f=0.5$. (d) Hysteresis loops for different system sizes $V$ at fixed $f=0.5$.}
\label{fig: combine_frequency}
\end{figure*}

For the driven Schl\"ogl network, the ensemble-averaged response $\langle x(t)\rangle$  traces out the hysteresis loop against the periodic drive $\epsilon_0\cos(2\pi f t)$. Individual stochastic trajectories do not show a clean periodic response because transitions between low- and high-concentration regimes occur at random times. By contrast, the ensemble average smooths out these trajectory-to-trajectory fluctuations and oscillates with the period of the external control, but with a phase lag, meaning that the response $\langle x(t)\rangle$ reaches its maxima and minima later than the external periodic drive $\epsilon_0\cos(2\pi f t)$, as shown in Fig.~\ref{fig: time_series}(c) and (d). This lag produces the closed response-drive loop shown in Fig.~\ref{fig: combine_frequency}(a)-(d).

This delay in the response originates from the competition between two fundamental time scales: the driving period of the external control and the intrinsic relaxation time of the system~\cite{das2013chaos, chen2025coercivity}. The resulting hysteresis loops are illustrated in Fig.~\ref{fig: combine_frequency}(a) for representative driving frequencies. Their shape, size, and symmetry depend strongly on the modulation frequency.

At low modulation frequency, for example $f=0.3$, the external control varies slowly enough that the system has sufficient time to follow the driving signal. Consequently, the response follows nearly the same path during the two halves of the driving cycle, producing a narrow hysteresis loop with a small phase lag.

At high modulation frequencies, for example $f=1.5$--$1.9$, the external control changes too rapidly for the system to respond within a driving period. In this regime, the dynamics are dominated more strongly by the intrinsic nonlinear relaxation and fluctuations than by the instantaneous value of the drive. The response therefore traverses similar paths during the two halves of the cycle, and the hysteresis loop shrinks.

At intermediate modulation frequencies, for example $f=0.9$--$1.1$, the driving period and intrinsic relaxation time are comparable. The response then partially follows the external drive while retaining a substantial phase lag, producing the largest hysteretic response.

We also examined how the hysteresis loop changes with parameters other than driving frequency: the well width $w$, driving amplitude $\epsilon_0$, and system size $V$. These parameters control distinct aspects of the response, including the effective landscape, forcing strength, and intrinsic fluctuation scale.

The well-width parameter $w$ changes both the effective bistable landscape and the strength of intrinsic fluctuations. The magnitude of $w$ is inversely proportional to the intrinsic noise amplitude: increasing $w$ corresponds to a lower level of fluctuations.~\cite{leonard1994stochastic}  Fig.~\ref{fig: combine_frequency}(b) shows hysteresis loops for representative values of $w$ at fixed driving frequency. At small $w$ ($0.5$--$0.6$), the loops are narrow, indicating a weak nonlinear response. As $w$ increases ($0.8$--$0.9$), the loops become broader and more elliptical, indicating a larger phase lag between the drive and the response. The nonmonotonic variation of loop width with $w$, increasing up to $w=0.8$--$0.9$ and then decreasing, indicates an optimal well width, or equivalently an optimal fluctuation regime, for dynamic hysteresis.

The driving amplitude $\epsilon_0$ controls the strength of the external modulation. Fig.~\ref{fig: combine_frequency}(c) shows hysteresis loops for representative amplitudes $\epsilon_{0}=0.05$--$0.15$ at fixed frequency $f=0.5$. Across this range, the loops remain centered near the same response value, while the loop width increases monotonically with $\epsilon_0$. At low amplitude ($\epsilon_0=0.05$--$0.07$), the system shows a smaller intrawell response to the periodic drive, producing a narrow loop. At larger amplitudes, the stronger drive enhances the intrawell oscillation amplitude and phase lag of the ensemble-averaged response, even though transitions to the high-concentration well remain limited by the state-dependent fluctuations.

System size $V$ controls the scale of intrinsic fluctuations. Fig.~\ref{fig: combine_frequency}(d) shows hysteresis loops for $V=1.001$--$1.401$ at fixed frequency $f=0.5$. In this model, physically meaningful parameter values require $V$ to remain above a lower threshold: for $V<1.001$, one or more of the rates $\alpha$, $\beta$, and $\gamma$ become negative, and the low-concentration minimum moves into an inaccessible region associated with negative control parameters. For $V>1.001$, the rates remain positive, and increasing $V$ reduces the strength of intrinsic fluctuations. As a result, noise-driven transitions become less frequent, the oscillation amplitude of the ensemble-averaged response decreases, and the hysteresis loop shrinks. Together, these trends show that the hysteretic response is controlled by both the external driving protocol and parameters that set the intrinsic fluctuation scale.

\subsection{Turnover in hysteresis loop area}
\label{subsec: Area}
The preceding subsection shows that the hysteresis loops change shape and size upon varying the driving and system parameters. To quantify these changes, we compute the area enclosed by the response-drive loop over one period of the external modulation. For a parameter $s\in\{f,\epsilon_0,V,w\}$, the hysteresis-loop area is
\begin{align}
    A_{\mathrm{hys}}(s) = \oint \langle x(t) \rangle\, d\epsilon,
    \label{eq:hys_area}
\end{align}
where $\epsilon(t)=\epsilon_0\cos(2\pi f t)$ is the periodic drive and $d\epsilon$ denotes the infinitesimal change in the drive $\epsilon(t)$ i.e., the differential of $\epsilon_0\cos(2\pi f t)$. Larger loop areas correspond to stronger hysteretic response and, in many driven systems, greater hysteresis loss. 

\begin{figure*}[ht]

\centering
    
    \includegraphics[width=0.48\linewidth, height=5.5cm]{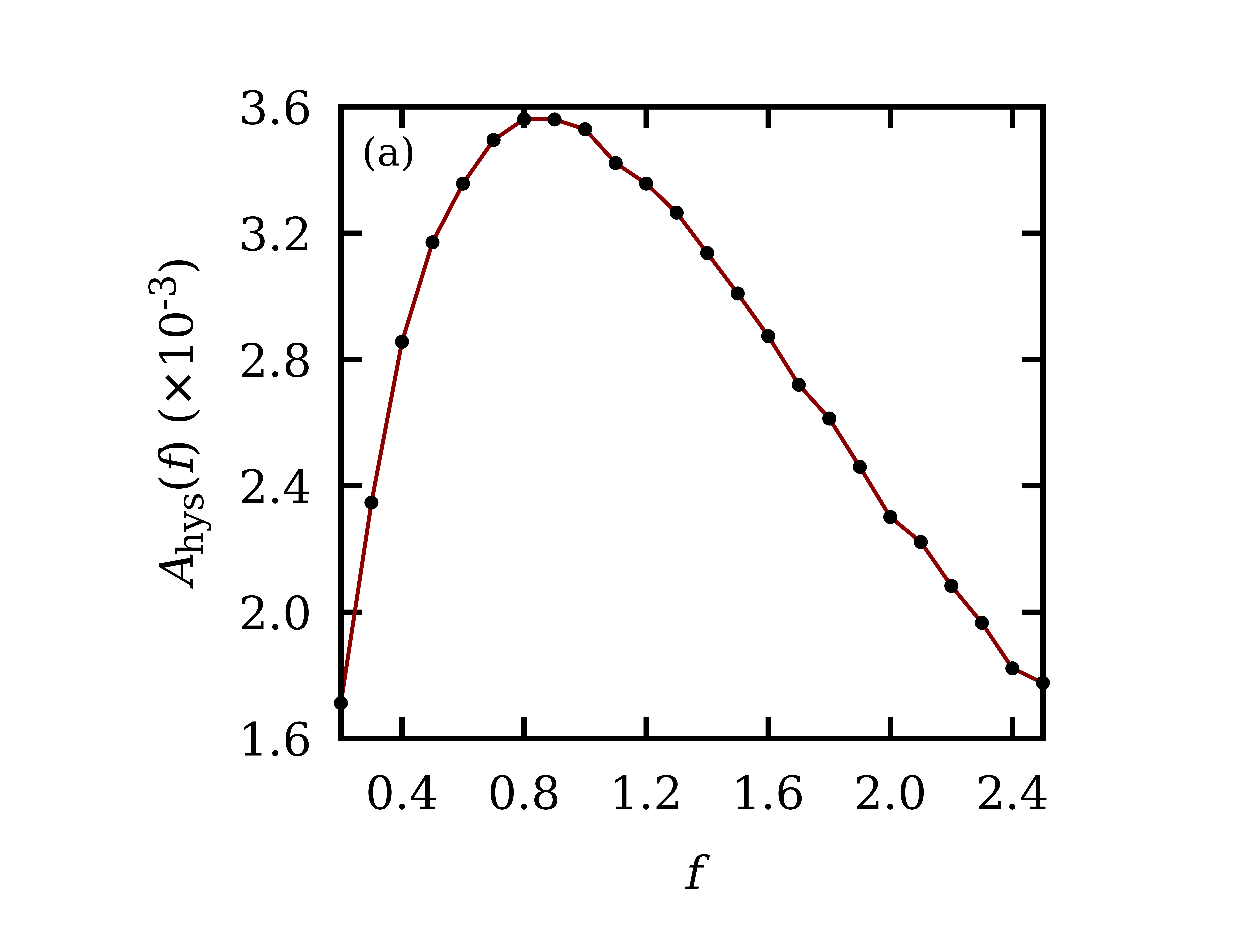}
    \includegraphics[width=0.48\linewidth, height=5.5cm]{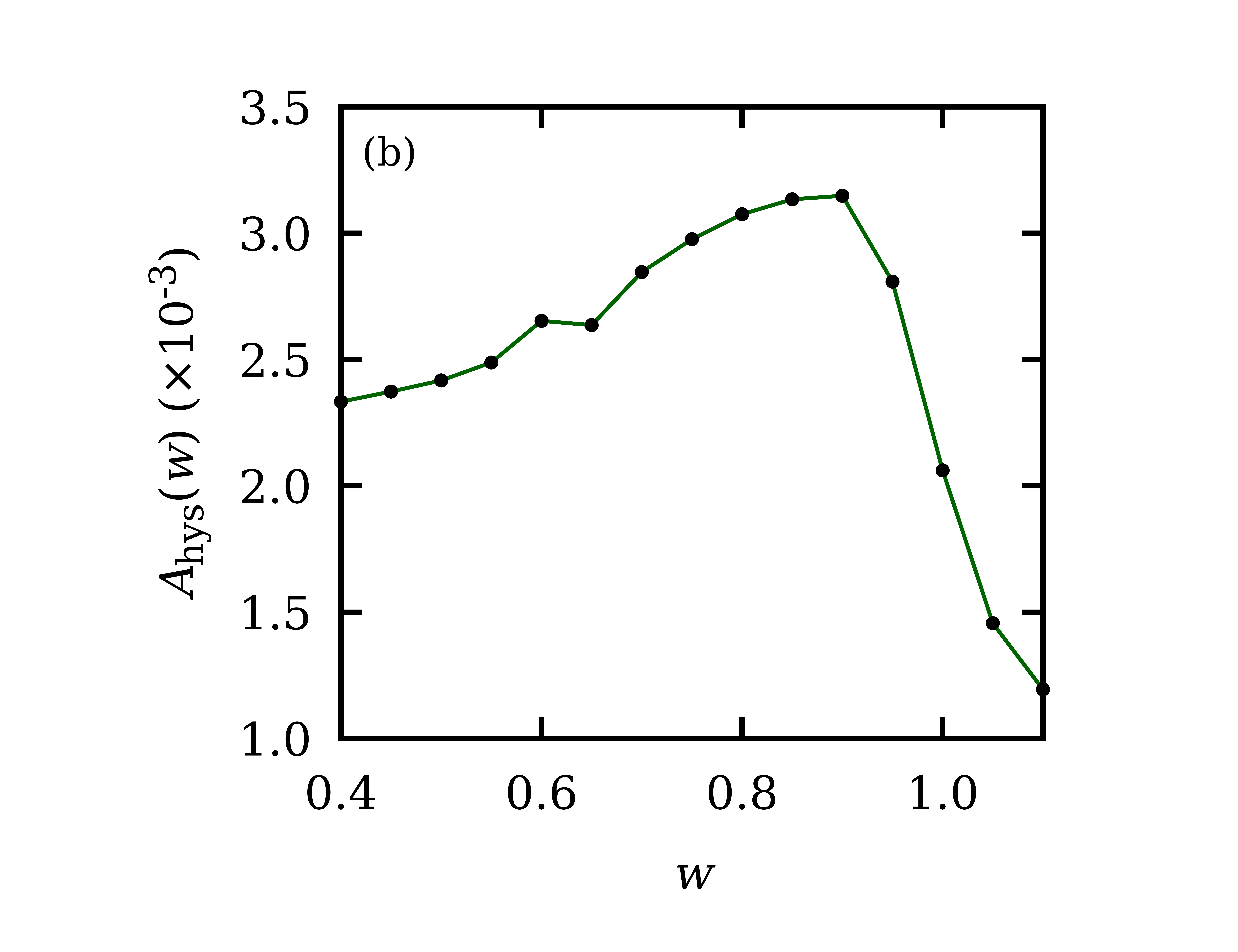}

\vspace{0.25cm}
  
    \centering

    \includegraphics[width=0.48\linewidth, height=5.5cm]{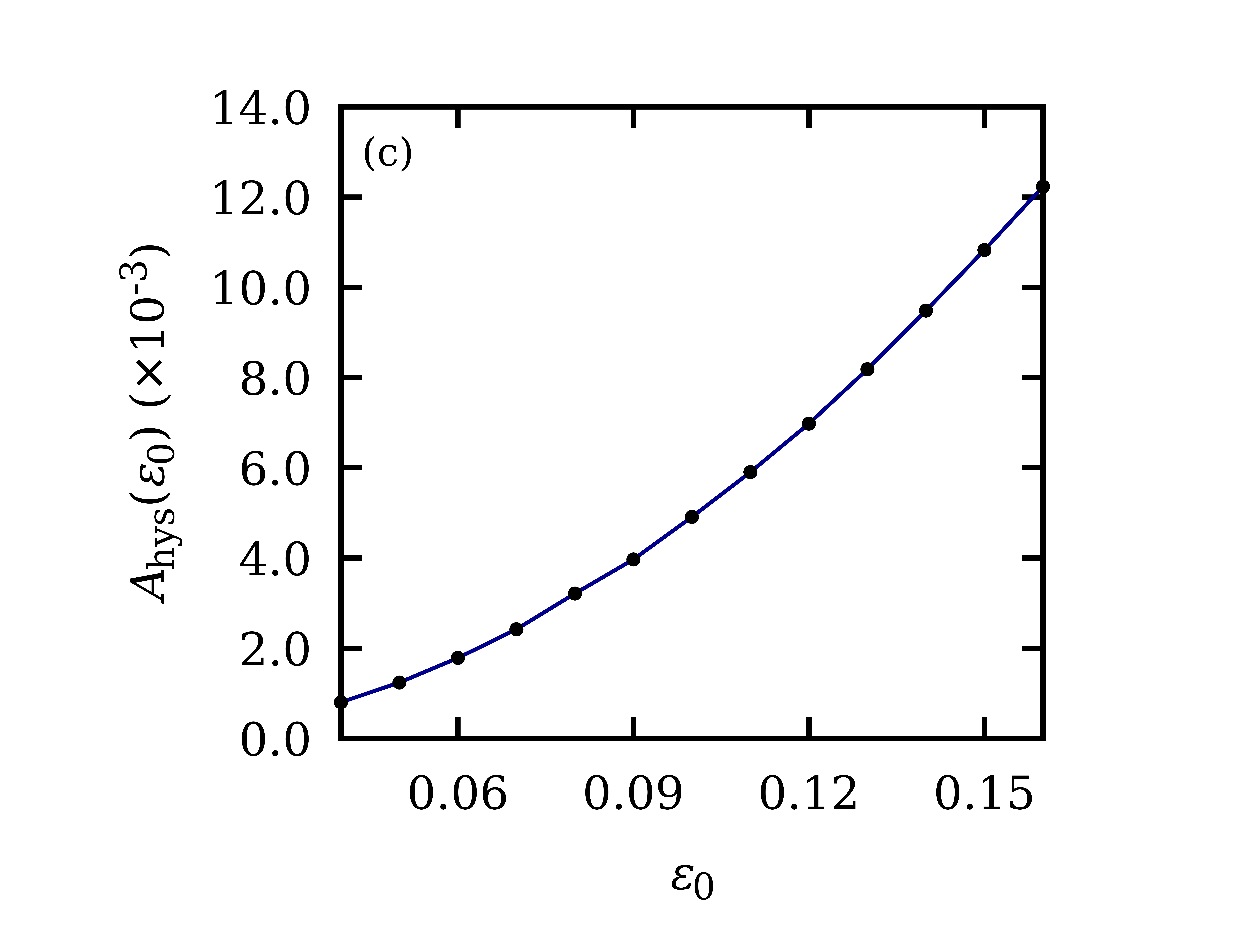}
    \includegraphics[width=0.48\linewidth, height=5.5cm]{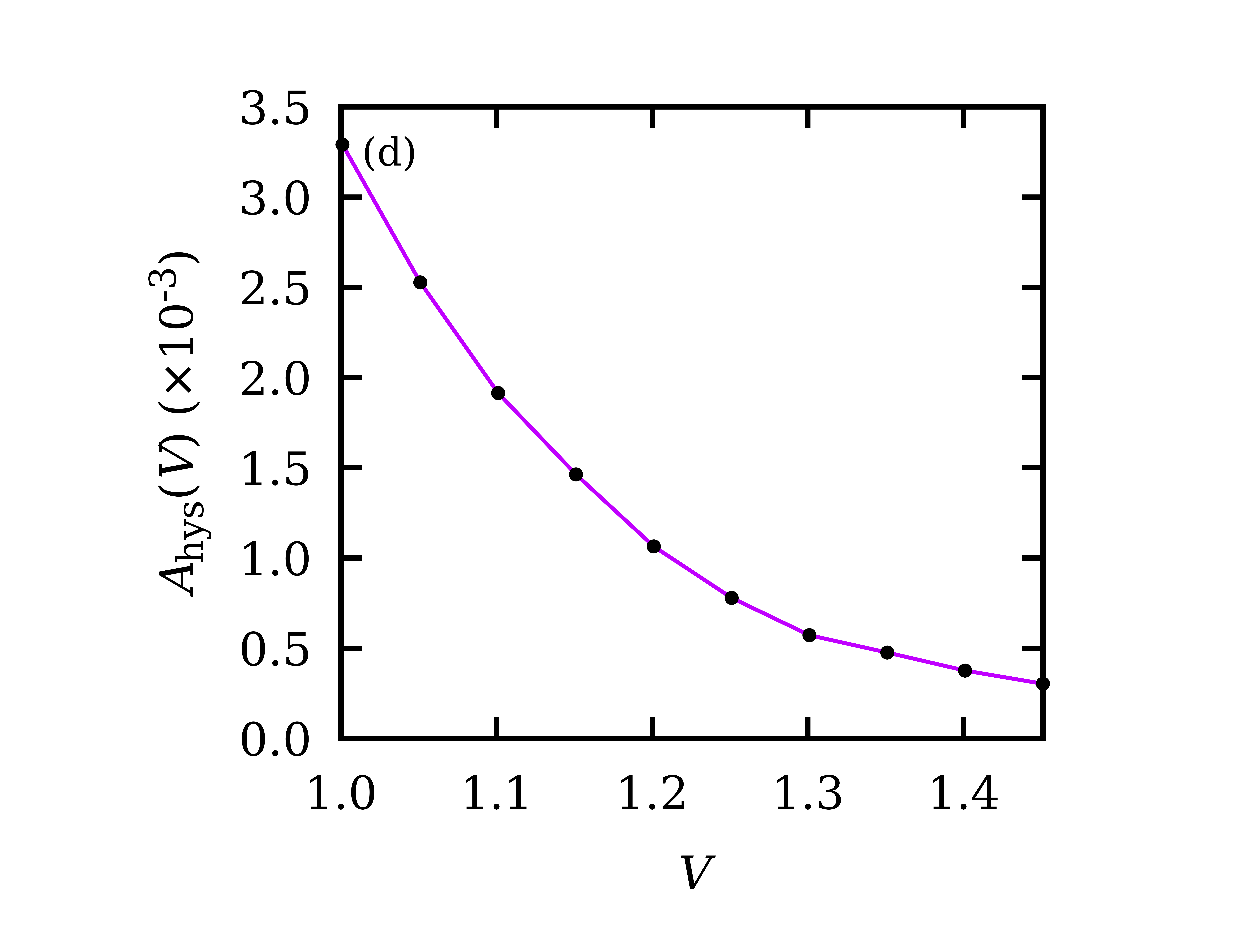}

      \caption{Parameter dependence of the dynamic hysteresis loop area. (a) Loop area $A_{\mathrm{hys}}(f)$ as a function of driving frequency showing a maximum near $f=0.8$--$0.9$. (b) Loop area $A_{\mathrm{hys}}(w)$ as a function of well width showing a maximum near $w=0.9$. (c) Loop area $A_{\mathrm{hys}}(\epsilon_0)$ as a function of driving amplitude showing a monotonic increase. (d) Loop area $A_{\mathrm{hys}}(V)$ as a function of system size at fixed $f=0.5$ showing a monotonic decrease.}  
        \label{fig: Area}
\end{figure*}

Initially, the loop area increases with driving frequency, but it eventually reaches a maximum and decreases at high frequencies~\cite{das2013chaos}. This turnover, shown in Fig.~\ref{fig: Area}(a), reflects the same time-scale competition discussed above: at low frequency the response nearly tracks the drive, whereas at high frequency the system cannot relax within a driving period. The maximum at intermediate frequency indicates the regime in which the delayed response is largest. The small loop area in the low-frequency limit is consistent with the dynamic nature of the hysteresis, which vanishes in the quasistatic limit. 

The loop area also varies nonmonotonically with the well width $w$, as shown in Fig.~\ref{fig: Area}(b). At small and large values of $w$, the hysteresis loop area is reduced, whereas an intermediate well width produces a maximum. This turnover is consistent with the trends in Fig.~\ref{fig: combine_frequency}(b): the hysteretic response is strongest when the effective landscape and intrinsic fluctuation scale produce the largest delayed response to the periodic drive.


The turnover of the hysteresis loop area at an intermediate value of the parameter $w$ representative of the fluctuation strength resembles the effect of stochastic resonance. 
As in stochastic resonance, the peak in loop area caused by dynamical hysteresis appears to be a consequence of the external drive having a period that matches of the system's intrinsic relaxation time scale. 
While these two dynamic processes have a similar origin, the scope of dynamic hysteresis is broader. Stochastic resonance primarily concerns the amplification of the weak periodic signal in the presence of fluctuations. However, dynamic hysteresis is not restricted by the magnitude of the external periodic signal.

In contrast to the frequency and well-width dependence, the hysteresis loop area increases monotonically with driving amplitude $\epsilon_0$, as shown in Fig.~\ref{fig: Area}(c). Increasing $\epsilon_0$ strengthens the periodic modulation and increases the amplitude of the intrawell response, thereby increasing the area enclosed by the response-drive loop.

The hysteresis loop area decreases monotonically with system size $V$, as shown in Fig.~\ref{fig: Area}(d). Because the intrinsic fluctuation strength decreases as $V$ increases, larger systems exhibit fewer noise-driven excursions and a smaller ensemble-averaged oscillation amplitude. The reduced response amplitude leads to a smaller response-drive loop area.


Together, these area trends show that dynamic hysteresis in the driven Schl\"ogl network is controlled by both the external protocol and the parameters that set the bistable landscape and fluctuation scale.

\section{Stochastic thermodynamic analysis}
\label{sec: stochastic thermodynamic analysis}




So far, our analysis suggests that chemical reactions can exhibit dynamical hysteresis and that the intrinsic kinetic time scales determine its extent. 
Now we turn to a thermodynamic analysis of this dynamical phenomenon. A particularly relevant thermodynamic quantity is the entropy of the system, which we might na\"ively expect to show a memory effect because of its relation to the information content of probability distributions. Given that hysteresis is typically associated with nonequilibrium processes, we also analyzed the entropy production rate~\cite{Pal2017,Zheng2026,Lucarini2010}. 

While we have analyzed dynamic hysteresis in the driven autocatalytic reaction system primarily through the Langevin dynamics framework, we complement that kinetic description through two approaches in which we quantify the the Shannon entropy of the system over a complete cycle of the external periodic drive. In one approach, we coarse-grain the dynamics into two states corresponding to the low- and high-concentration regimes of the autocatalyst. In a second approach, we calculate the Shannon entropy directly from the solution of the chemical master equation. 
We find that the Shannon entropy measured through both of these methods exhibits 
dynamic hysteresis under external driving. These hysteresis and their loop area show behavior similar to the concentration response function $\langle x(t) \rangle$, including a turnover at nearly the same intermediate frequency.  
This comparison allows us to assess how the information-theoretic measure of Shannon entropy reflects the process of dynamic hysteresis.
To characterize the thermodynamic irreversibility associated with dynamic hysteresis, we also evaluate the entropy production rate over a complete period of the external drive using the chemical master equation. 

\begin{figure*}[ht]

\centering
    
    \includegraphics[width=0.49\linewidth, height=5.5cm]{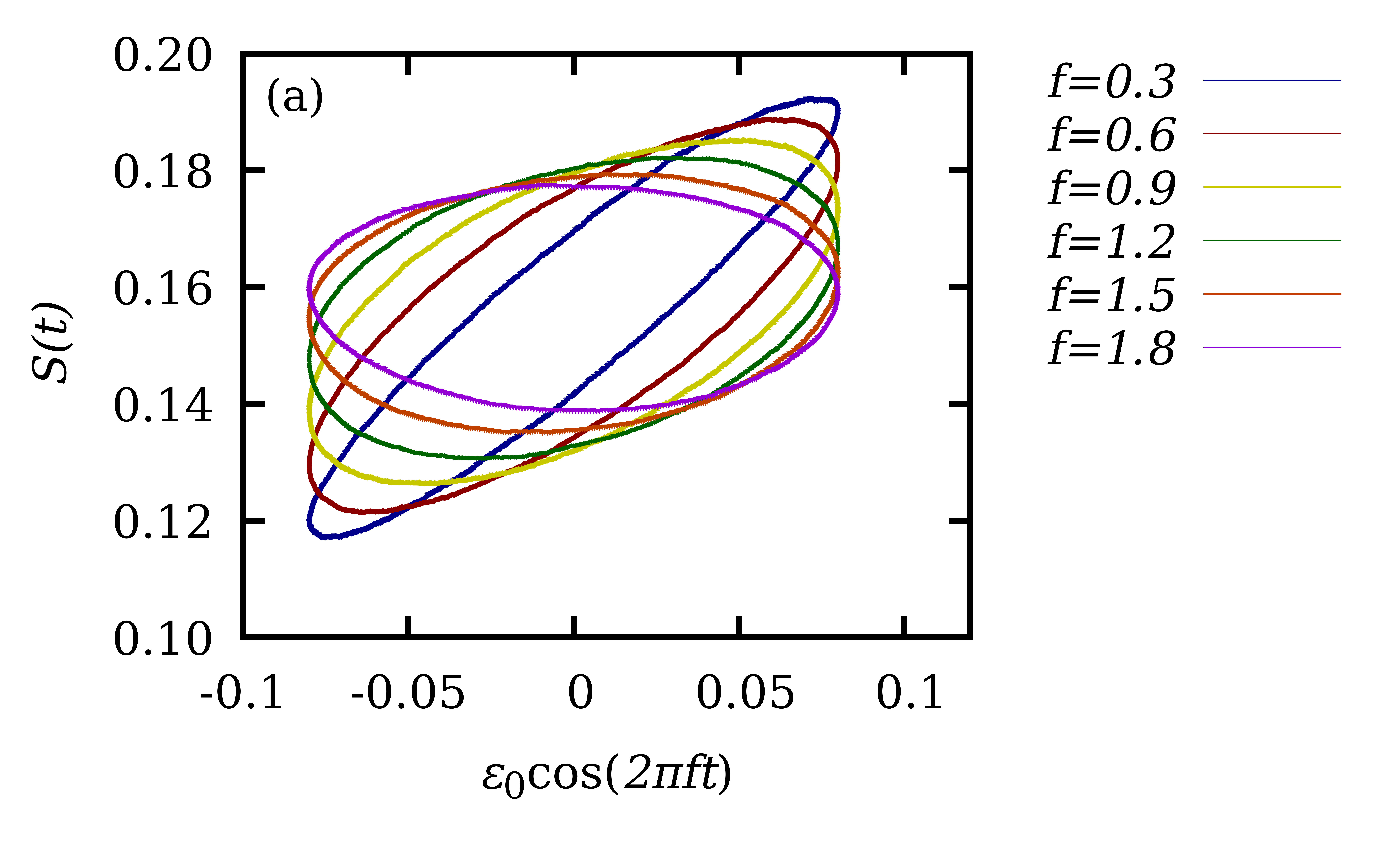}
    \includegraphics[width=0.49\linewidth, height=5.5cm]{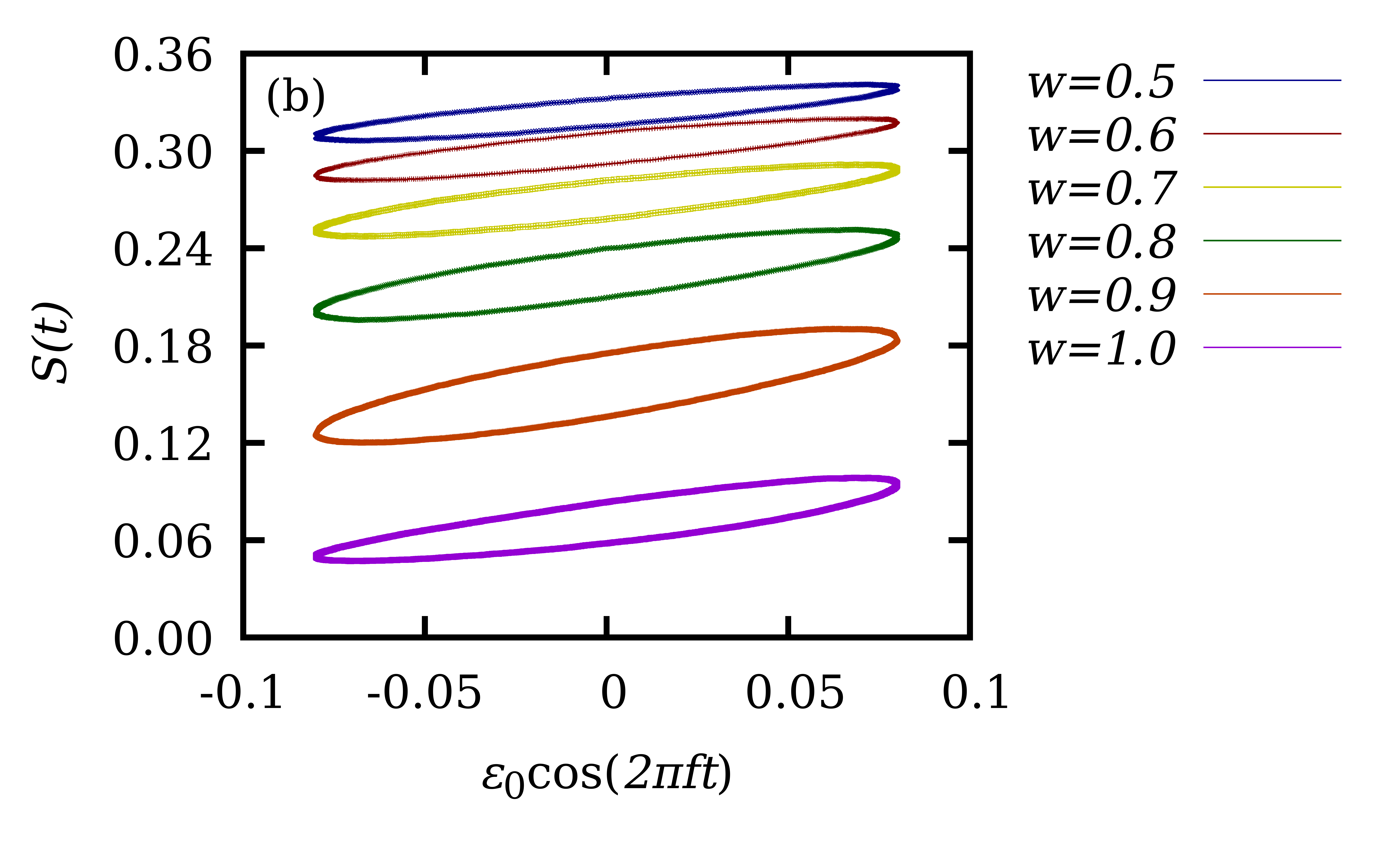}

\vspace{0.25cm}
  
    \centering

    \includegraphics[width=0.49\linewidth, height=5.5cm]{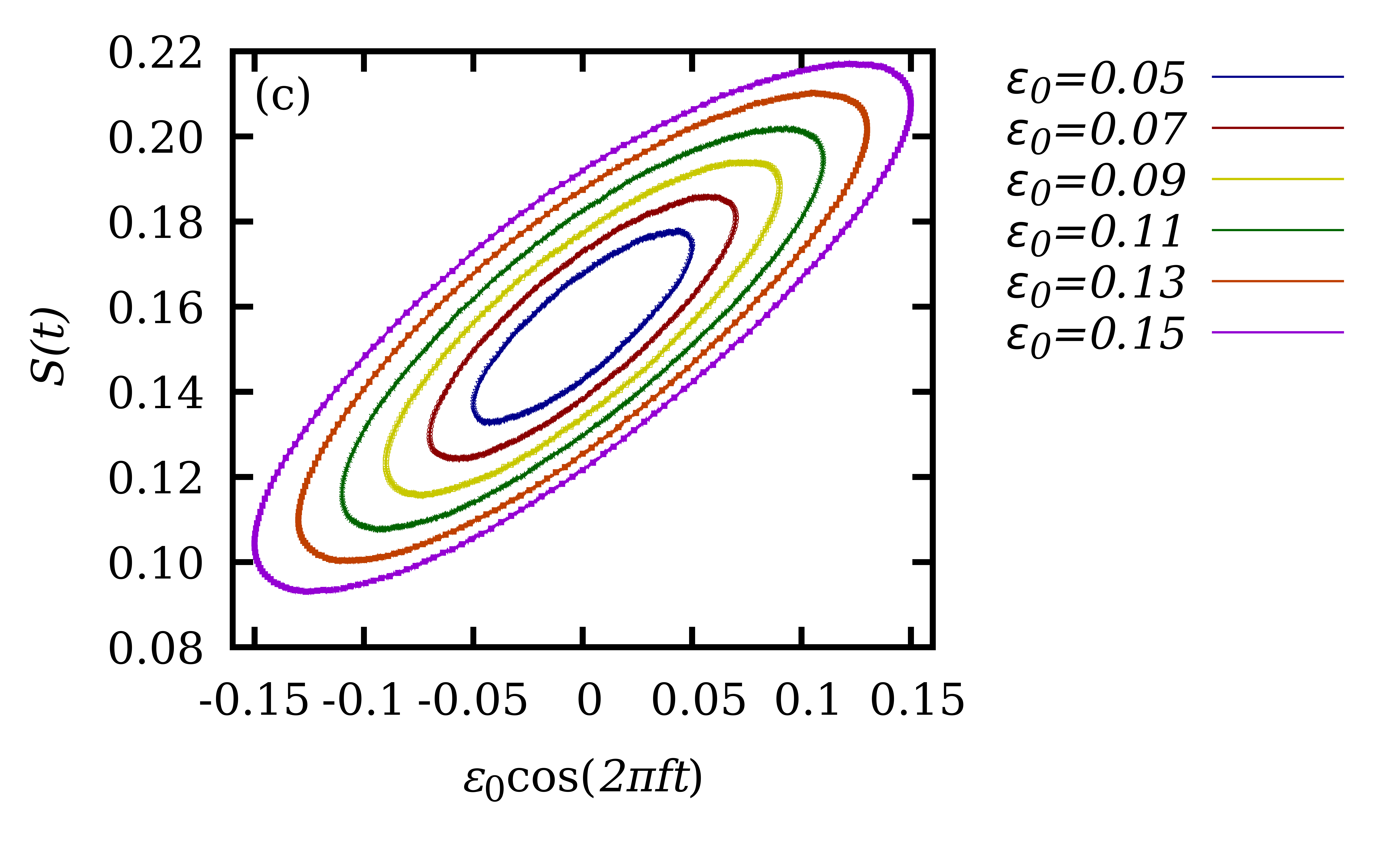}
    \includegraphics[width=0.49\linewidth, height=5.5cm]{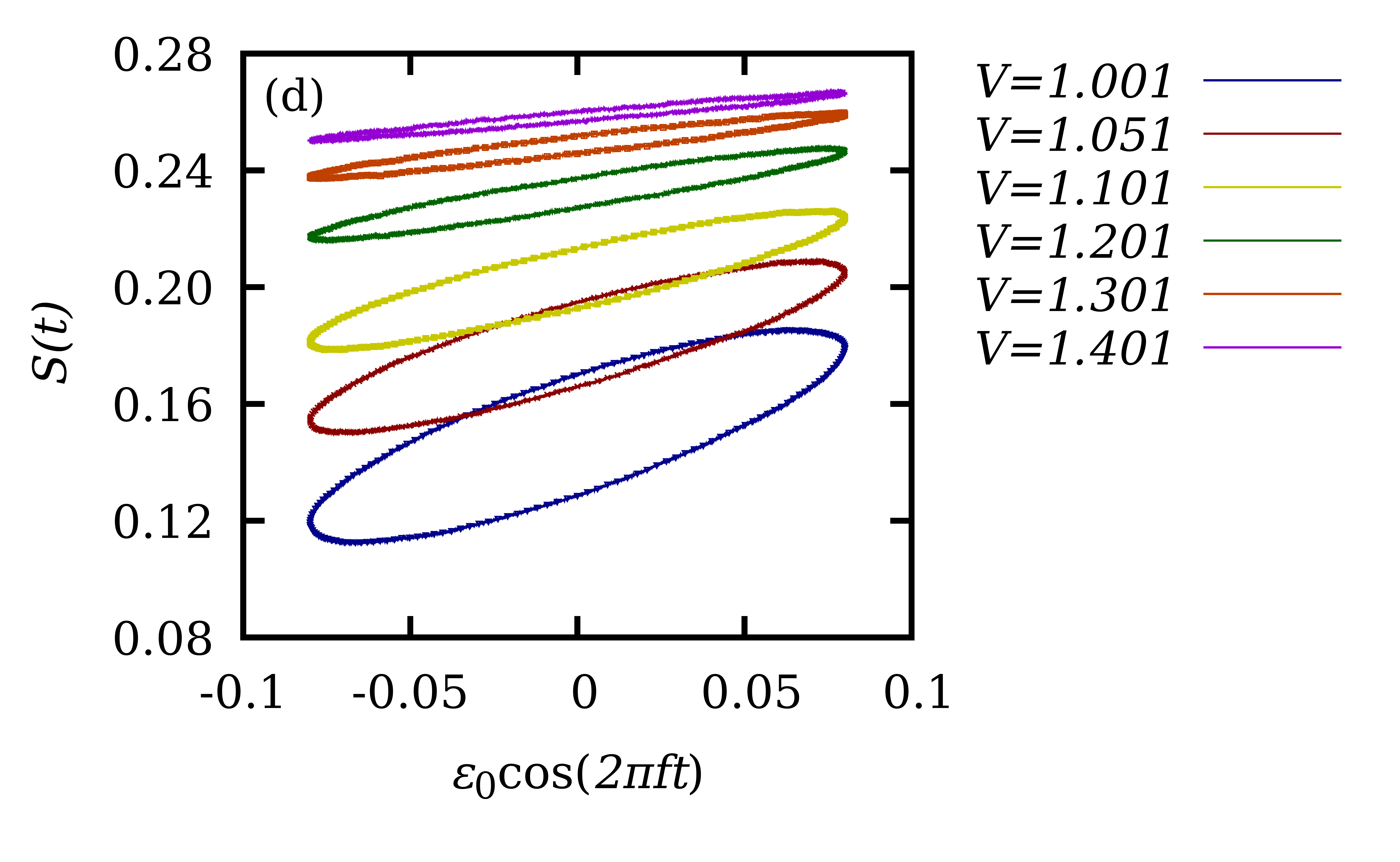}

        \caption{Hysteresis loops representing the uncertainty in the concentration of the autocatalytic species $X$, quantified via information entropy $S(t)$ as a function of external periodic force $\epsilon_{0} \cos(2\pi f t)$. (a) Loops for different frequencies $f$ of driving periodic force. (b) loops at a fixed driving frequency $f = 0.5$, showing the effect of varying the potential well width $w$. (c) Loops for varying the amplitude of the periodic force $\epsilon_{0}$ at fixed frequency $f = 0.5$. (d) Loops showing the response of the system under the effect of varying system size $V$ at fixed frequency $f = 0.5$}
        \label{fig: S_hys}
\end{figure*}
\subsection{Shannon entropy: Two-state model}
\label{subsec: Shannon entropy}

Shannon entropy is a measure of the system's information content and state uncertainty, and hysteresis is related to information storage and information loss. Therefore, we might then anticipate that the system's information entropy can be considered as an important observable in analyzing dynamical hysteresis. The Shannon entropy can be calculated at the trajectory level~\cite{seifert2005entropy} for the Langevin dynamics. 
The Langevin description incorporates the notion of a deterministic gradient of an effective double-well potential along with the state-dependent intrinsic fluctuations (Eq.~(\ref{eq: stochastic_diff_eq})). The bistability of this underlying double-well potential generates two concentration regimes for the autocatalytic species. 
In the present chemical system, dynamic hysteresis is governed by transitions between these two distinct concentration zones, where the left and right regions of the double-well structure, representing the low- and high-concentration states, correspond to $x(t) < x_{0}$ and $x(t) > x_{0}$, respectively.
The dynamic hysteresis in this context involves transitions between two states under periodic control as the basic mechanism. Therefore, we consider a coarse-graining based on the two distinct states of the system. 

To calculate the Shannon entropy we use the residence probabilities for the two specific states, $L$ and $R$, corresponding to the low- and high-concentration regions of the autocatalytic species. 
By definition, the residence probabilities are
\begin{equation}\label{eq:occupation_prob_LR}
P_{L}(t)=\int_{0}^{x_{0}} p(x,t)\,dx,
\qquad
P_{R}(t)=\int_{x_{0}}^{\infty} p(x,t)\,dx.
\end{equation} 
The lower limit of $x$ to calculate $P_{L}(t)$ can be zero for all possible concentrations to be physically meaningful. The upper limit of $x$ for the estimation of $P_{R}(t)$ can mathematically be $\infty$ according to the underlying bistable structure of the potential. However, that is not physically consistent in reality. It has a finite value that is determined by the periodic drive and the intrinsic fluctuations. However, to calculate $P_{L}(t)$ and $P_{R}(t)$ practically through numerical simulations, the lower limit of $x$ for $P_{L}(t)$ and the upper limit of $x$ for $P_{R}(t)$ need not be given explicitly. We examine whether $x$ is less than or greater than $x_{0}$. The stochastic trajectories at time $t$ corresponding to the former case contribute to $P_{L}(t)$, and those satisfying the latter condition at any arbitrary time $t$ contribute to $P_{R}(t)$. Therefore, $P_{L}(t)$ and $P_{R}(t)$ represent the integrated probabilities of the low and high concentration zones, respectively, of the autocatalyst concentration.

Now, we define the Shannon entropy of the system by taking into account the low and high autocatalyst concentration regions as the two definite states of the system. 
The instantaneous Shannon entropy for the system is specified with the following formula,
\begin{eqnarray}\label{eq:shannon_eq}
S(t) &=& -\sum_{i = L, R}P_{i}(t) \ln P_{i}(t) 
\end{eqnarray}
The Shannon entropy, defined for the system in this way, provides its value for a two-state model of the reaction network under consideration. 
To examine how the system's state uncertainty varies under periodic control, we study the quantity $S(t)$ as a function of the periodic drive. Similar to the response function, the Shannon entropy exhibits hysteresis with respect to the external periodic control. We analyzed this hysteresis behavior for the relevant controlling parameters, such as frequency $f$ and amplitude $\epsilon_{0}$ of the periodic drive, fluctuation strength $w$, and system volume $V$. 

Fig.~\ref{fig: S_hys}(a) illustrates the dynamic hysteresis loops of the Shannon entropy $S(t)$ as a function of external periodic drive in the driving frequency range of $0.3-1.8$. 
In the low-frequency regime, the hysteresis loop is narrow, showing that the probabilistic response of the system closely follows the external drive, with a small lag between them. The system has enough time to relax in the potential wells, allowing occasional transitions between them. As a result, $S(t)$ spans a wider range during the hysteresis cycle, exhibiting both low- and high-entropy states. However, the average Shannon entropy $S(t)$ over a cycle generally attains lower values in this regime, indicating comparatively ordered states.  
At intermediate frequencies, the hysteresis loops become broader, 
indicating synchronized transitions between wells. This behavior arises from the matching of the intrinsic relaxation time scale of the system with that of the external drive. Consequently, the probabilistic spread over available states is enhanced and retained almost over the whole cycle. Therefore, the hysteresis loop area attains high values in this frequency range.
At much higher frequencies, the hysteresis loops shrink 
as the system cannot respond effectively to the rapidly oscillating external driving force. 
This mainly leads to oscillations of $S(t)$ around its average value, thereby reducing the effect of hysteresis.   
The trend of the variation of the features of the hysteresis loops for $S(t)$ with changing $f$ matches that for $\langle x(t) \rangle$. 

The time evolution of the system subject to the modulation of the external drive, for parametric change in the well-width parameter $w$, representing a measure of the fluctuation strength, is demonstrated by Fig.~\ref{fig: S_hys}(b). This illustrates the dynamic hysteresis loops for different values of $w$. The results indicate that for smaller values of $w$, for which the effective strength of the intrinsic noise is higher, there is an increase in the Shannon entropy of the system. This suggests that a higher level of randomness in the system, emerging due to the greater extent of fluctuations, corresponds to a larger entropy value for it. In this range, the hysteresis loops are narrow, as the dynamics are mostly dominated by the intrinsic fluctuations. As a result, the response and consequently the Shannon entropy of the system do not experience the effect of the external periodic field substantially to exhibit a significant hysteretic effect. As $w$ increases, the effective noise intensity decreases, keeping the Shannon entropy value for the system in a comparatively smaller range, with reduced uncertainty in the dynamics. In this range of $w$, the intrinsic relaxation time scale of the system and the time-period of oscillation of the external control become comparable. Therefore, the entropy response of the system exhibits a proper lag with the external driving, which leads to broader hysteresis loops. 
Further increase in $w$ reduces the strength of the fluctuations notably. In this case, the entropy response of the system is primarily controlled by the extrinsic periodic modulation. Consequently, the former is able to closely follow the latter. This results in smaller hysteresis loops. 
The characteristic change of the hysteresis loops for $S(t)$ corresponding to the parametric variation of $w$ is similar to that for the response function $\langle x(t)\rangle$. 

The effect of varying the amplitude of the periodic drive is demonstrated by the dynamic 
hysteresis loops for $S(t)$ shown in Fig.~\ref{fig: S_hys}(c). 
The hysteresis loops for $S(t)$ for the range of 
$\epsilon_{0}$ studied, suggest that $S(t)$ spans to a 
wider extent for the larger values of $\epsilon_{0}$. 
This follows from the greater degree of oscillations of the 
response function $\langle x(t) \rangle$ with increasing 
amplitude $\epsilon_{0}$ of the periodic control. 
Consequently, the parametric change of $\epsilon_{0}$ has a similar effect on the hysteresis loops for $S(t)$ as that for $\langle x(t) \rangle$. 

Next, we examine the dependence of the Shannon entropy $S(t)$ on the external periodic 
drive subject to the parametric change in the volume of the system $V$. The resulting behavior is illustrated in the form of dynamic hysteresis loops, as shown in Fig.~\ref{fig: S_hys}(d). 
The hysteresis loops show that as we decrease the volume of the system, the size of the loops becomes larger. 
It implies that when the volume of the system is small, large intrinsic fluctuations arise. 
They assist the system in overcoming the barrier between the two concentration zones of the species $\ce{X}$, facilitating transitions between the wells by amplifying the weak periodic signal. 
The increase in the absolute Shannon entropy value with an increase in volume is a 
consequence of the greater available physical space for the species to explore. 
This is consistent with the extensivity property of the entropy, in general, i.e., 
entropy is proportional to the system volume.
Similar to the other parameters, $V$ has the same effect on the features of the hysteresis loops for $S(t)$ and $\langle x(t) \rangle$. 

\begin{figure*}[ht]

\centering
    
    \includegraphics[width=0.49\linewidth, height=5.5cm]{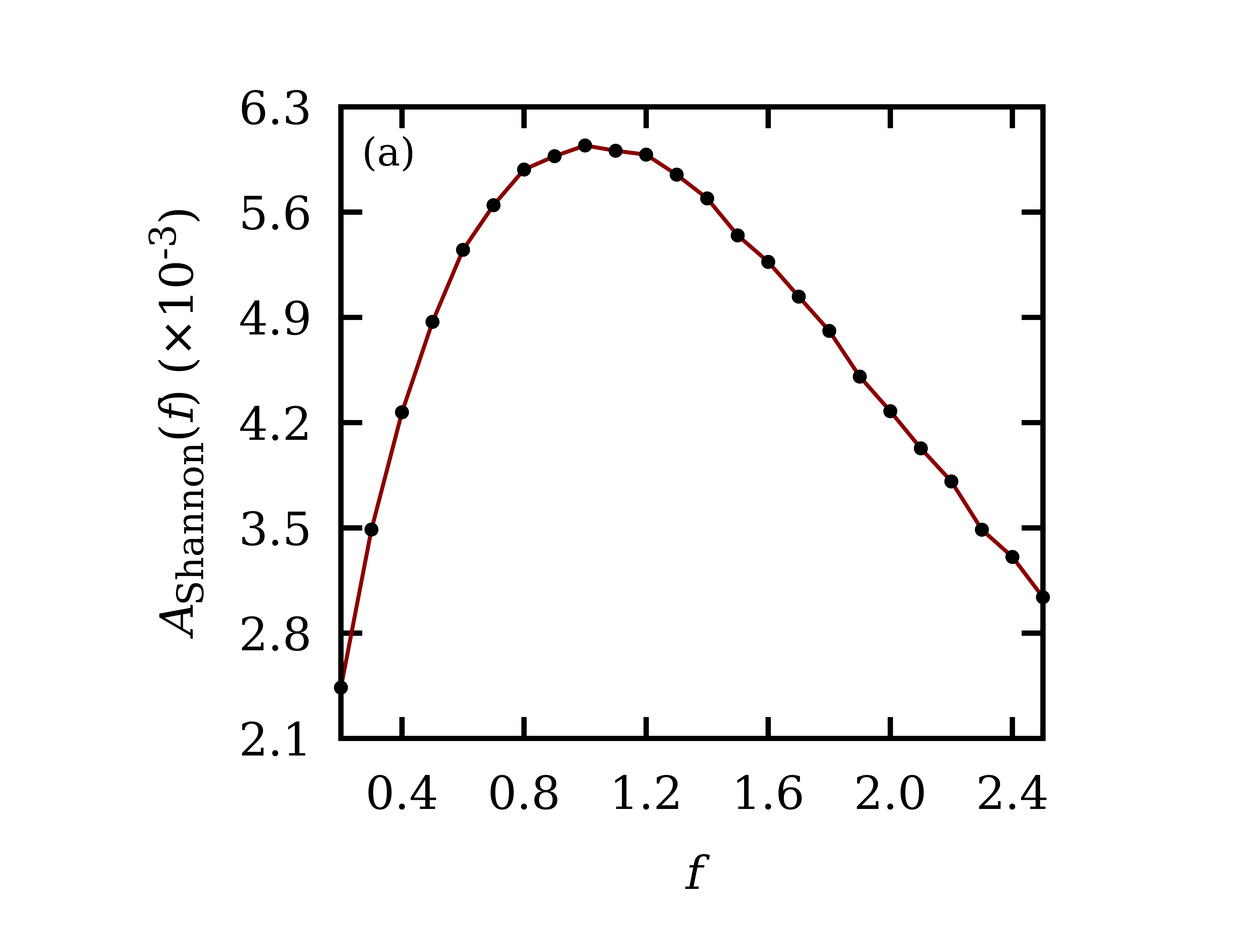}
    \includegraphics[width=0.49\linewidth, height=5.5cm]{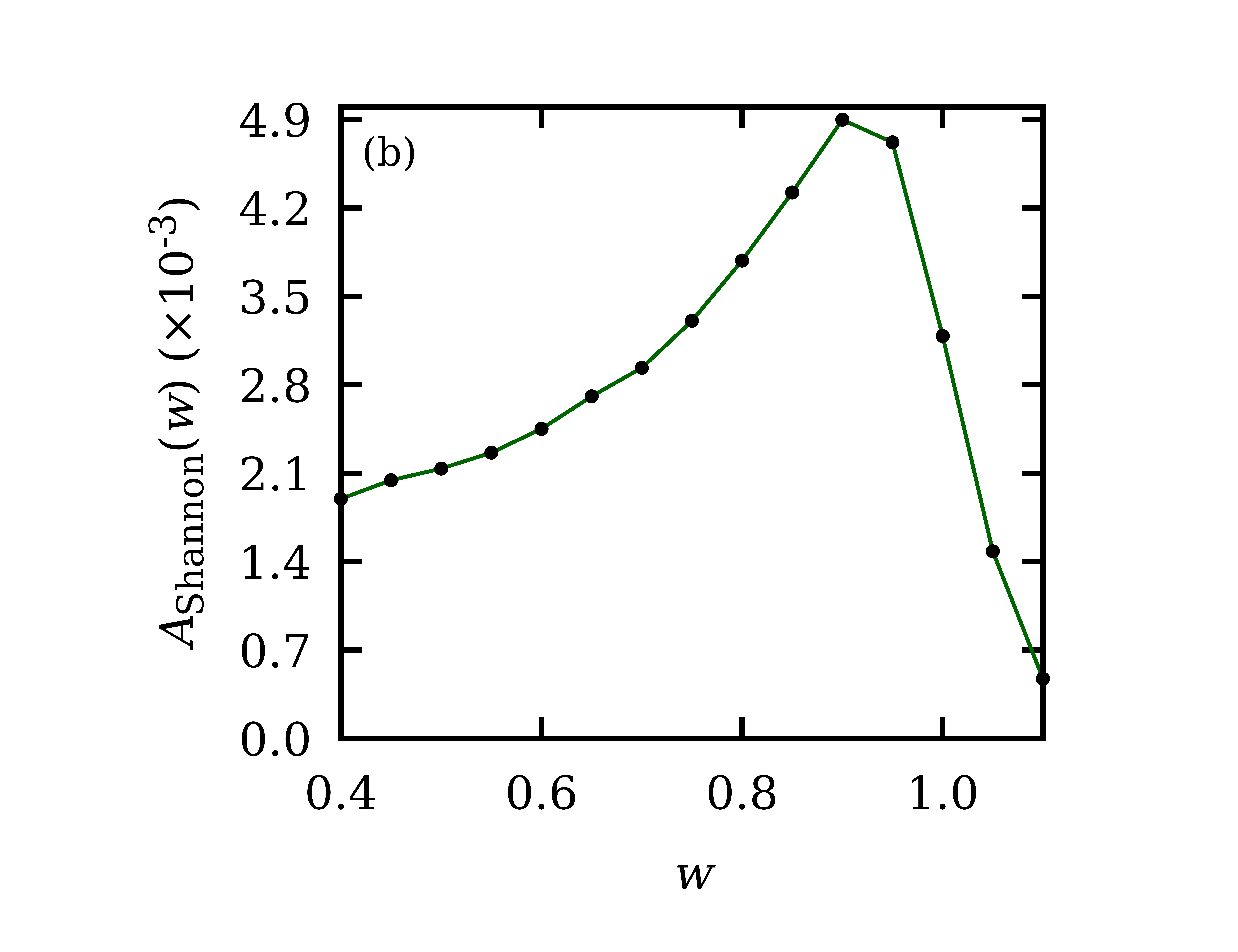}

\vspace{0.25cm}
  
    \centering

    \includegraphics[width=0.49\linewidth, height=5.5cm]{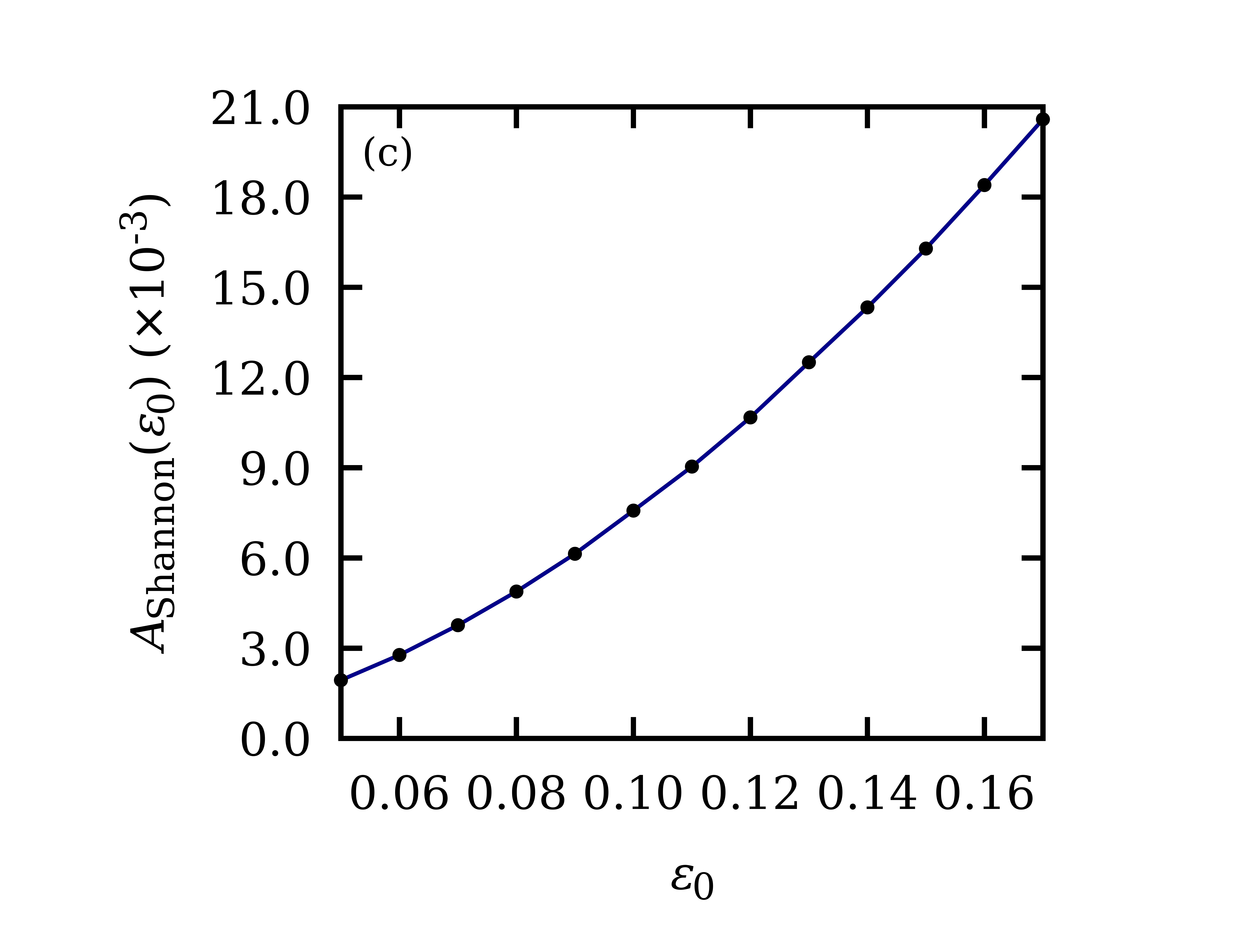}
    \includegraphics[width=0.49\linewidth, height=5.5cm]{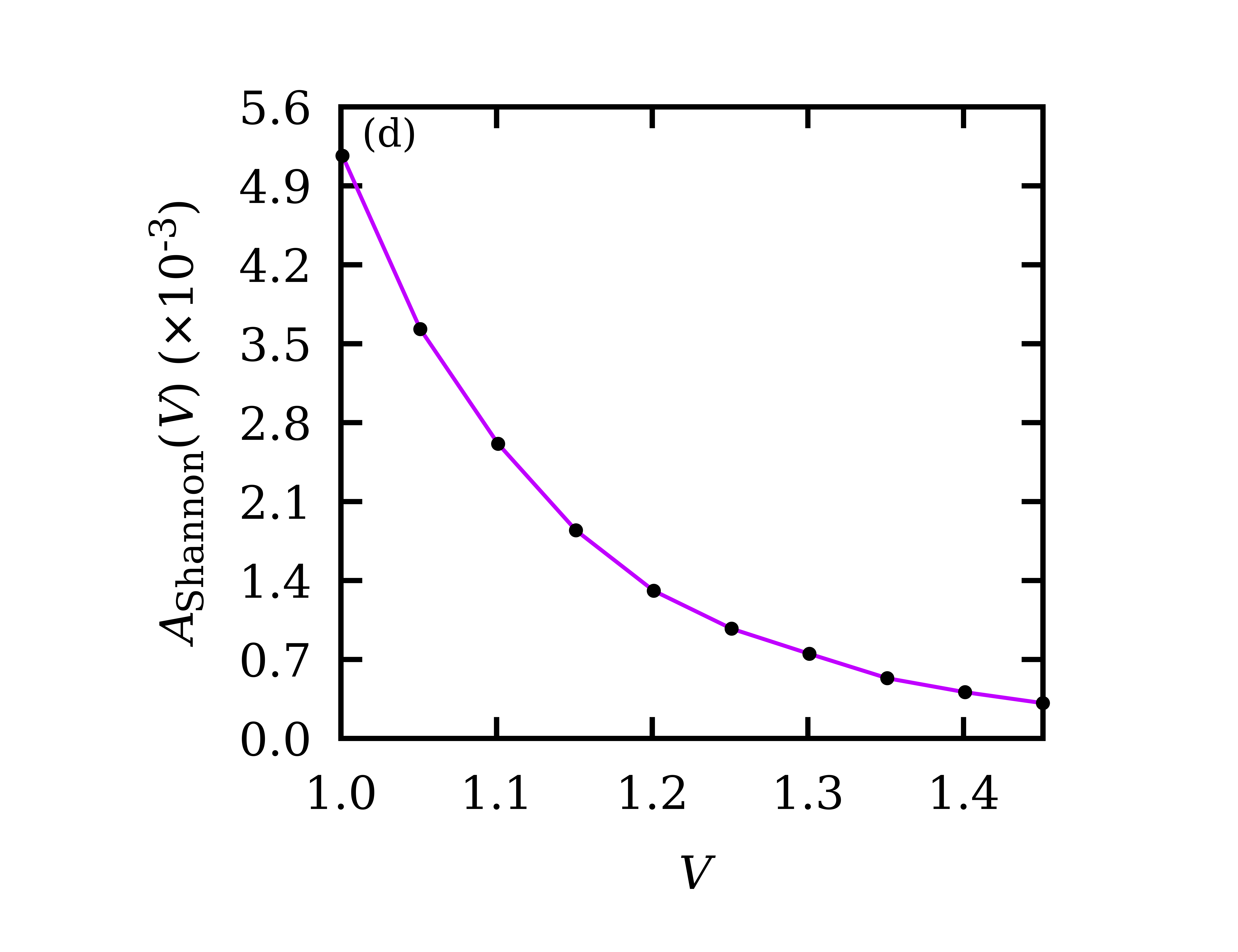}

        \caption{(a) Hysteresis loop area as a function of driving frequency $f$; the maximum hysteresis loop area occurs at $f = 1.0$. (b) Effect of width of the potential-well on hysteresis loop area with a maximum at $w = 0.9$. (c) Monotonic increase in hysteresis loop area as a function of the amplitude $\epsilon_{0}$ of external periodic force. (d) Effect of system size $V$ on the hysteresis loop area showing a monotonic decrease.}
        \label{fig: S_hys_area}
\end{figure*}

\subsection*{Loop area turnover}
Next, we aim to understand the variation of the attributes of the hysteresis loops with respect to the changes in the controlling parameters more quantitatively. For this purpose, we study the hysteresis loop area, estimating the extent of hysteresis, in respective situations. It is calculated numerically by integrating the value of $S(t)$ over a complete period, over the change of the external periodic drive $\epsilon(t) = \epsilon_{0} \cos ({2 \pi f t})$. The area covered by the hysteresis loop is defined as\\
\begin{equation}
    A_{\text{Shannon}} =  \oint \langle S(t) \rangle d\epsilon
    \label{eq:S_hys_loop_area}
\end{equation} 
The hysteresis loop area for $S(t)$ quantifies the irreversibility of information involvement in one cycle with respect to the variation of the extrinsic periodic control. The quantity is studied against the change of the relevant governing parameters: frequency $(f)$ and amplitude ($\epsilon_{0}$) of the external periodic drive, fluctuation strength $(w)$, and volume of the system $(V)$. The results have been illustrated in Figs.~\ref{fig: S_hys_area}(a)-(d). 

The hysteresis loop area is very small at the low and high frequencies, and has a maximum value at an intermediate frequency. This has been illustrated in Fig.~\ref{fig: S_hys_area}(a). This observation matches the trend of the corresponding results for $\langle x(t) \rangle$. Importantly, the loop area turnover frequency ($f = 1.0$) for $S(t)$ coincides with that for $\langle x(t) \rangle$ (Fig.~\ref{fig: Area}(a)). These similar behaviors suggest that the dynamic response and the evolution of the probability distribution in the accessible states, and eventually the Shannon entropy, are strongly coupled and follow analogous kinetics. 

The hysteresis loop area variation as a function of the fluctuation strength $(w)$ is shown in Fig.~\ref{fig: S_hys_area}(b). We see that at lower and higher values of the potential width, the hysteresis loop area is very small. Similar to the case for the variation of $f$, the maximum in the hysteresis loop area is observed at an intermediate value of the controlling parameter $w$. That peak at $w = 0.9$ demonstrates maximum hysteretic response of the system in terms of the involvement of the Shannon entropy. As discussed above, the same behavior is detected for the $\langle x(t) \rangle$ - $\epsilon(t)$ hysteresis loop area (Fig.~\ref{fig: Area}(b)), having turnover and maximum response at $w = 0.9$. This close resemblance in the observations for $S(t)$ and $\langle x(t) \rangle$ in the present context again highlights their immediate connection. We suggest it as a link between a kinetic observable, the response function $\langle x(t) \rangle$, and an information-thermodynamic quantity $S(t)$ for the process of dynamic hysteresis.   

After this, we calculate the hysteresis loop area as a function of the amplitude of the periodic force $\epsilon_{0}$ and the system volume $V$. We observe that the loop area increases monotonically with increasing amplitude of the external modulation  ($\epsilon_{0} = 0.05 - 0.17$), while it decreases steadily with increasing system volume ($V = 1.001-1.501$). The results have been presented in Figs.~\ref{fig: S_hys_area}(c) and (d) for these two cases, respectively. An increase in hysteresis loop area for $S(t)$ as we increase the amplitude $\epsilon_{0}$ occurs due to the elevated extent of oscillation amplitude of the system's response, encompassing a larger span of probability space. The size of the system directly affects the intrinsic fluctuations in the system. As we increase the volume of the system, it suppresses the noise effect, leading to very few transitions between wells. As a result, the system becomes less responsive towards periodic transitions caused by the external periodic driving. Consequently, the hysteresis loop area decreases with increasing system volume, shown in Fig.~\ref{fig: S_hys_area} (d). 

\subsection{Shannon entropy: Chemical master equation approach}
\label{subsec: Instantaneous Shannon entropy}
The dynamics of the autocatalytic species can also be described with the chemical master equation. In these kinetics, the instantaneous state of the system at time $t$ is defined by a probability mass function $P_{n}(t)$ that follows the conditions $P_{n}(t)\geq 0$ and $\sum_{n=0}^{n_{\max}} P_{n}(t) = 1$. Here, $n$ represents the number of the autocatalytic species $\ce{X}$ at any instant of time $t$. In this discrete Markovian system, the probability $P_{n}(t)$ evolves with time $t$ under stochastic transitions $t_{n}^+(t)$ and $t_{n}^-$ according to the chemical master equation Eq.~(\ref{eq:master_eq}) \cite{mcquarrie1967stochastic,vanKampen2007,Gardiner2009}. 

To ensure consistency with the underlying reaction stoichiometry and volume-dependent scaling of molecular interactions, the transition rates are derived using mass-action kinetics \cite{vellela2009stochastic}:
\begin{subequations}\label{eq:birth_death_rate}
\begin{align}
t_{n}^+(t) = k_{1}A \frac{n(n - 1)}{V} + k_{4}V C(t) \label{eq:birth_rate}\\
t_{n}^- = k_{2} \frac{n(n-1)(n-2)}{V^2} + k_{3} B n. \label{eq:death_rate}  
\end{align} 
\end{subequations}
Here, $k_{i}$ $(i = 1-4)$ are the stochastic reaction rate constants for the reactions occurring in a well-mixed vessel with volume $V$. The prefactors associated with identical reactant combinations $(k_{1}=k^{'}_{1}/2, k_{2}=k^{'}_{2}/6)$ are absorbed into the definitions of corresponding reaction rate constants. The concentrations of species $\ce{A}$, $\ce{B}$, and $\ce{C}$ are denoted by $A$, $B$, and $C$, respectively, and $C$ acts as the periodic driving source, with concentration
\begin{align}
     C(t)=C_{0}+C_{1}\cos(2 \pi f t)
\end{align}
For convenience, we define an effective time-dependent driving parameter:
\begin{align}
    \gamma(t) = k_{4}V C(t).
\end{align}
The numerical simulations are performed using the dimensional form of the chemical master equation with the transition probabilities described by Eqs~\ref{eq:birth_death_rate} (a) and (b). The transition rates $t_{n}^+(t)$ and $t_{n}^-$ are formulated with the dimension of inverse time, to maintain the dimensional consistency of the chemical master equation. The dimensions of the parameters are presented in Table~(\ref{tab:parameters}).

Shannon entropy measures the amount of randomness related to the microscopic state of the system $n$ at time $t$ in terms of probability distributions~\cite{shannon1948, Lesne2014,seifert2012}. For this description of the 
dynamics, the Shannon entropy is defined as,
\begin{equation}\label{eq:shannon_eq_CME}
S(t) = -\sum_{n}^{n_{\max}}P_{n}(t) \ln P_{n}(t). 
\end{equation}

In this study, we numerically solve the chemical master equation with the help of the second-order Runge-Kutta method with a constant time step $h = 10^{-4}$. 
The considered values of the rate parameters are presented in TABLE~(\ref{tab:parameters}), 
\begin{table*}[t]
\caption{Parameters used in the analysis of Shannon entropy and total entropy production rate. 
In the figures, we included numeric values of relevant parameters. Their dimensions are below. 
}
\label{tab:parameters}
\centering
\begin{tabular}{lccc}
\hline\hline
Parameter & Description & Dimension & Value \\
\hline
$V$ & System volume & [$L^{3}$] & 100 \\
$k_{1}$ & Rate constant & $[N^{-1}L^{6}T^{-1}]$ & 1.0 \\
$k_{2}$ & Rate constant & [$L^{6} T^{-1}$] & 1.0 \\
$k_{3}$ & Rate constant & [$N^{-1} L^{3} T^{-1} $] & 1.0 \\
$k_{4}$ & Rate constant & $[N^{-1}T^{-1}]$ & 2.0 \\
$A$ & Concentration of species A & $[N L^{-3}]$ & 0. 3 \\
$B$ & Concentration of species B & $[N L^{-3}]$  & 2.7 \\
$C_{0}$ & Mean driving concentration & $[N L^{-3}]$ & 1.0 \\
$C_{1}$ & Amplitude of external drive & $[N L^{-3}]$ & 0.08 \\
\hline\hline
\end{tabular}
\end{table*}
\begin{table*}[t]
\caption{Dimensions of variables used in stochastic thermodynamic analysis of chemical master equation}
\label{tab:quantities}
\centering
\begin{tabular}{lcc}
\hline\hline
Variable & Description & Dimension \\
\hline
$J_{n}(t)$ &Transition probability current & $[T^{-1}]$ \\
$A_{n}(t)$ & Affinity & Dimensionless \\
$\gamma (t)$ & Driving parameter & $[T^{-1}]$ \\
$S(t)$ & Shannon entropy & Dimensionless \\
$\dot{S}_{\text{total}}(t)$ & Total entropy production rate & $[T^{-1}]$ \\
$A_{\text{Shannon}}$ & Shannon entropy hysteresis loop area & $[T^{-1}]$ \\
$A_{\text{EPR}}$ & Entropy production hysteresis loop area & $[T^{-2}]$ \\
\hline\hline
\end{tabular}
\end{table*}

The simulation is performed for $100$ driving cycles with the maximum autocatalytic species number set to $400$. The system is driven by an external oscillating drive $\gamma(t)$. By solving the chemical master equation Eq.~(\ref{eq:master_eq}) with the above details, we obtain the probabilities needed to calculate the Shannon entropy according to Eq.~(\ref{eq:shannon_eq_CME}). 

\begin{figure*}[ht]

\centering
    
    \includegraphics[width=0.49\linewidth, height=5.5cm]{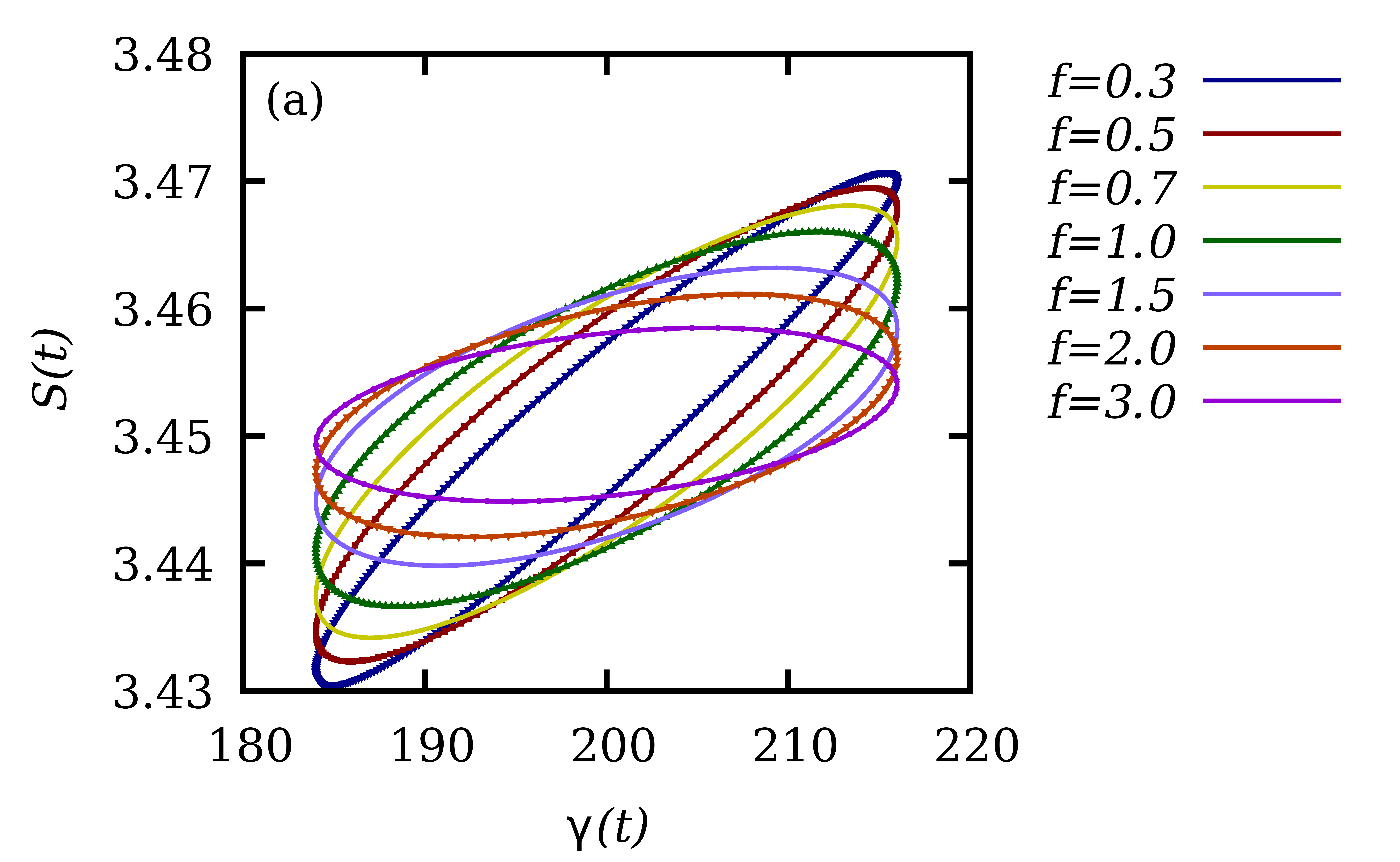}
    \includegraphics[width=0.49\linewidth, height=5.5cm]{Area_S_2k_2.7B.pdf}

        \caption{(a) Dynamic hysteresis loops for varying frequencies of driving force with respect to the Shannon entropy $S(t)$ and external periodic drive $\gamma(t)$, at different driving frequencies $f$. (b) Hysteresis loop area as a function of driving frequency $f$, showing maximum response at $f = 1.0$, illustrating the most nonequilibrium state of the system.}
        \label{fig:instantaneous_S}
\end{figure*}

We have illustrated the dynamic hysteresis loops for $S(t)$ obtained in this method in Fig.~\ref{fig:instantaneous_S}(a). The hysteresis loop area variation as a function of the driving frequency $f$ has been shown in Fig.~\ref{fig:instantaneous_S}(b).
These illustrations demonstrate that at low and high frequencies, the hysteresis loops are narrow. Consequently, the loop areas have smaller values. This implies that the system is not far away from equilibrium in these cases. However, at intermediate values of the frequency, the hysteresis loops become wider, covering more area, corresponding to highly nonequilibrium states.

In the two Shannon-entropy subsections above, we compare the hysteretic behavior of $S(t)$ obtained from the continuous (Langevin) and discrete (chemical master equation) descriptions of the present system. We then relate the hysteresis behavior of $S(t)$ in these two cases to that of the dynamical variable $\langle x(t) \rangle$ obtained from the Langevin equation. The dynamic hysteresis loops and area turnover obtained from both approaches for $S(t)$, and those estimated for $\langle x(t) \rangle$ within the Langevin framework, exhibit similar features, with maximum hysteretic response at frequency $f = 1.0$. These findings provide evidence for the consistency of stochastic and information-thermodynamic descriptions across different levels of system dynamics. 
The results also suggest the continuous Langevin is a good approximation of the discrete chemical master equation under these conditions.

\begin{figure*}[ht]

\centering

    \includegraphics[width=0.48\linewidth, height=5.5cm]{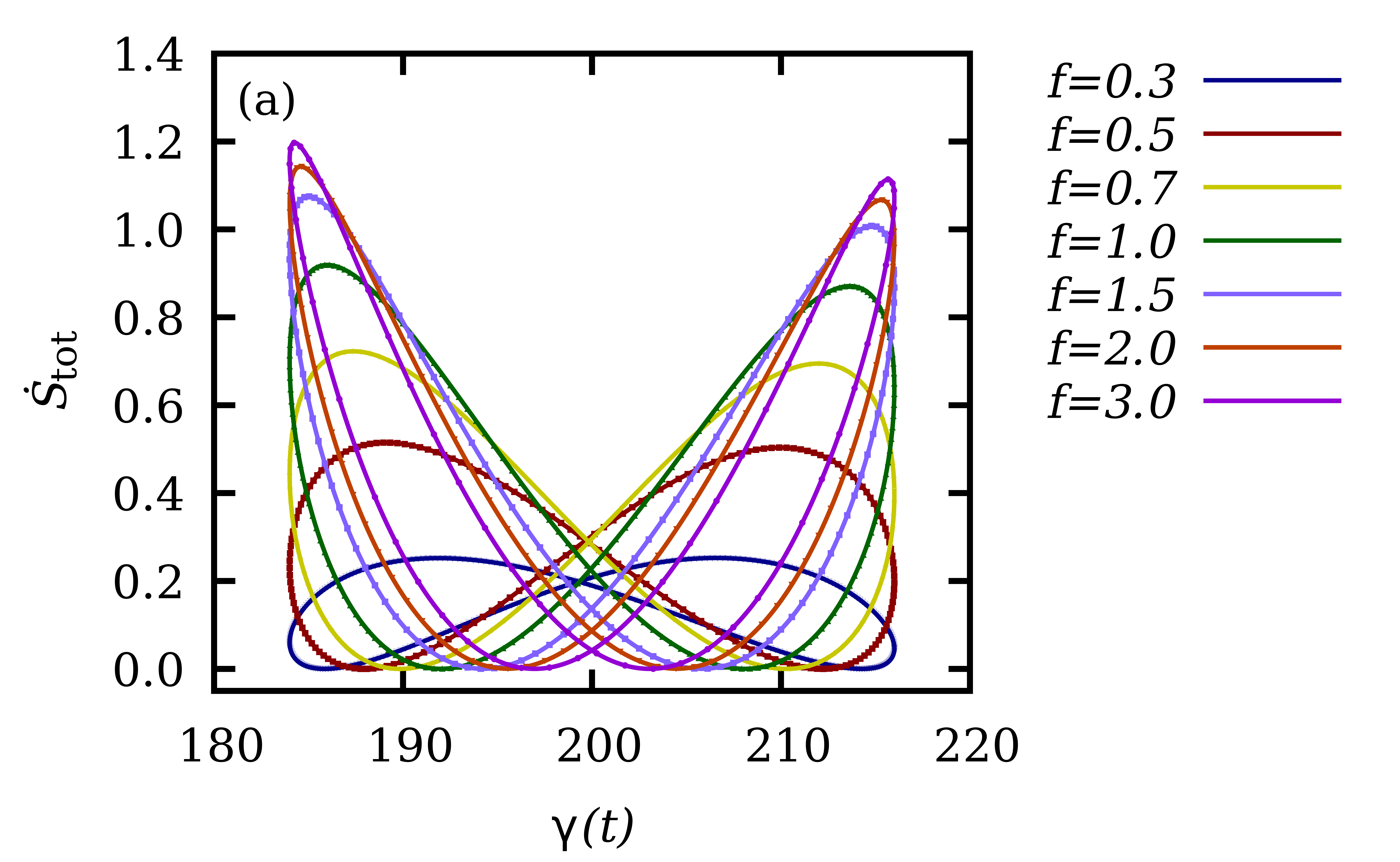}
    \includegraphics[width=0.48\linewidth, height=5.5cm]{EPR_J_A_Area_inter_2.7B_2k4.pdf}

        \caption{(a) Evolution of the butterfly-shaped hysteresis loop in the total entropy production rate $\dot{S}_{tot} (t)$ with changing frequency of the external time-dependent drive $\gamma(t)$. (b) The hysteresis loop area plot $A(f)$ versus $f$ indicates the dynamic nature of the loops and shows a maximum at frequency $f = 1.0$}
        \label{fig:instantaneous_EPR}
\end{figure*}

\begin{figure*}[ht]
        \centering

      \includegraphics[width=0.48\linewidth, height=5.5cm]{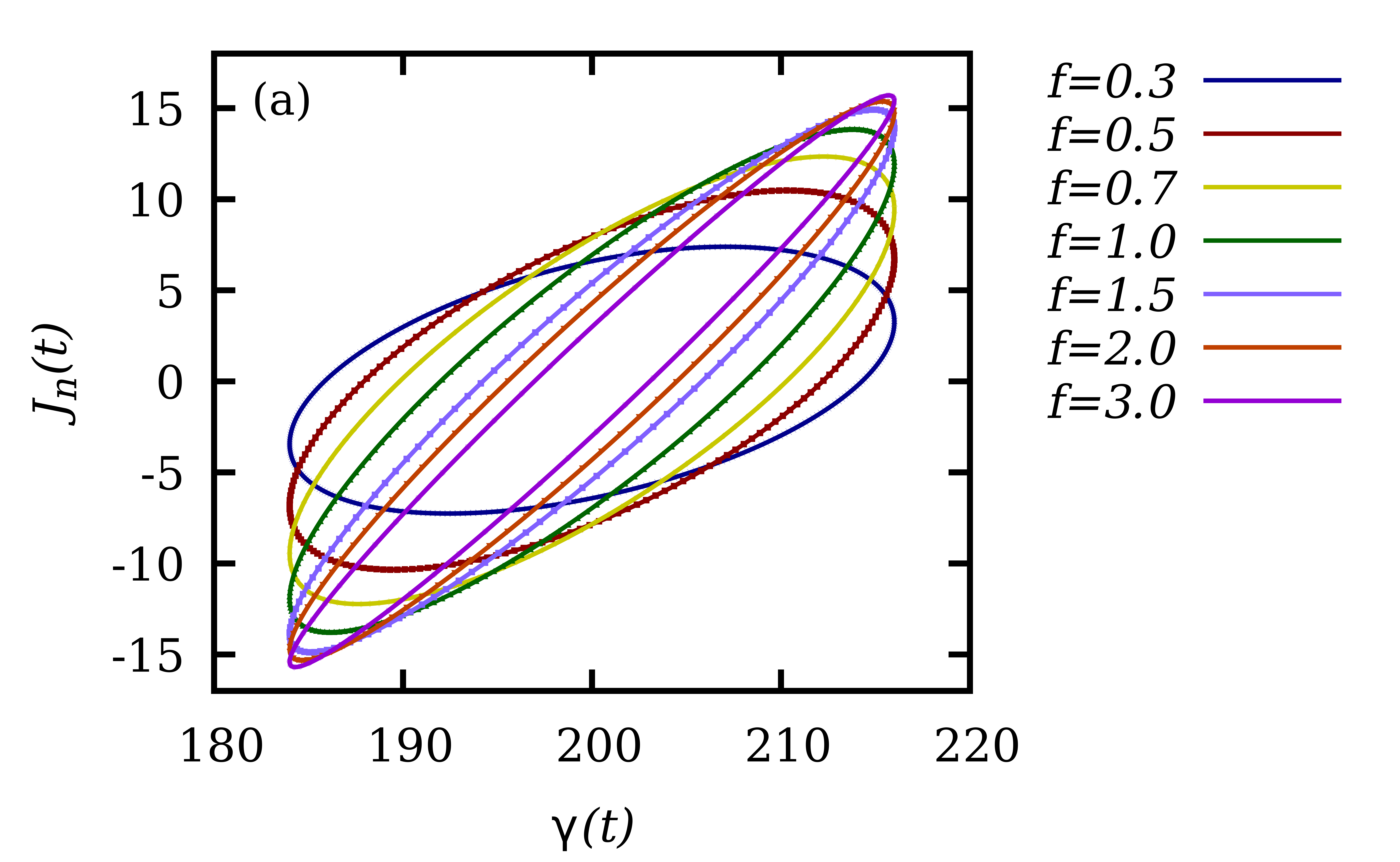}
      \includegraphics[width=0.48\linewidth, height=5.5cm]{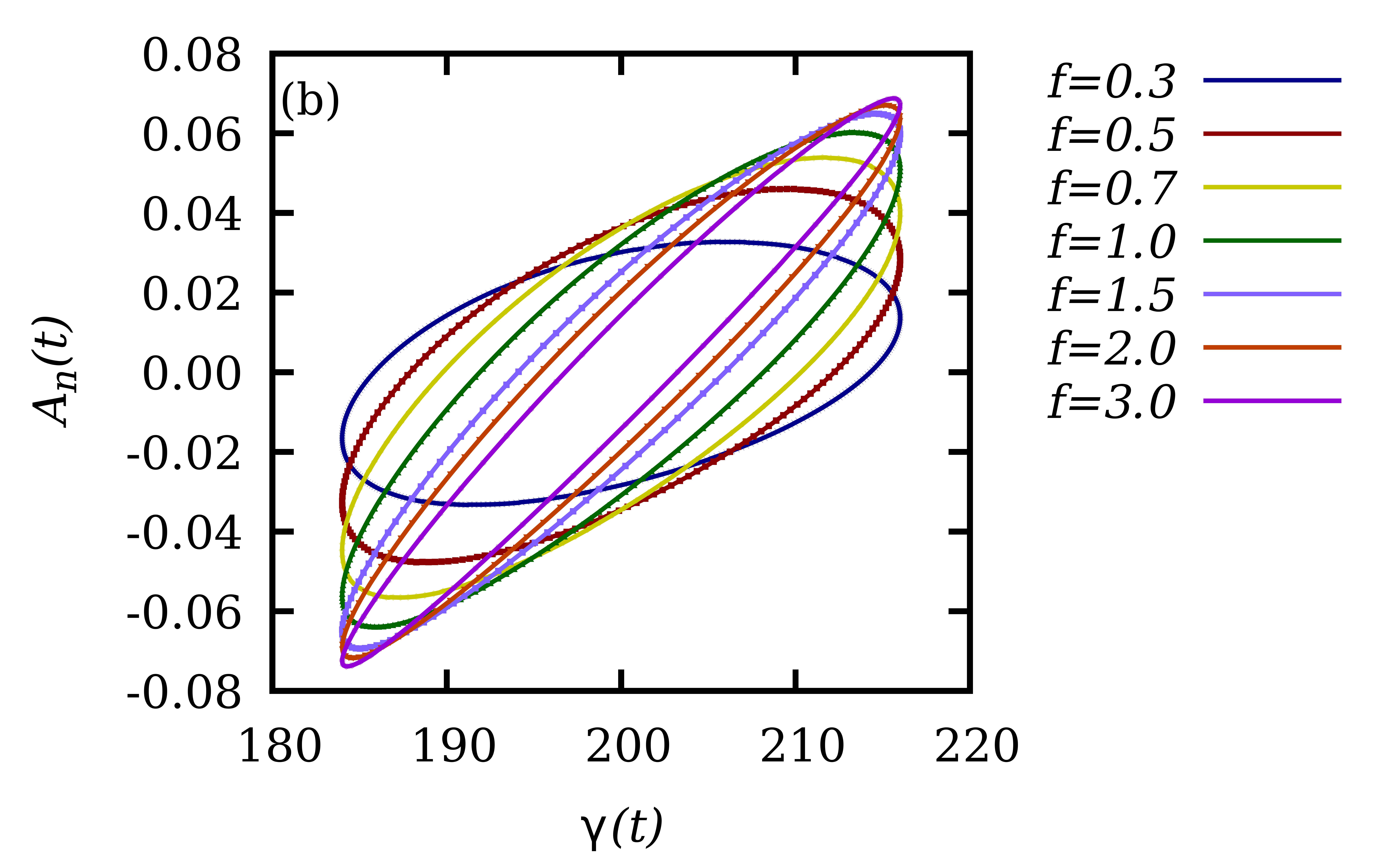}
      \includegraphics[width=0.6\linewidth, height=6.5cm]{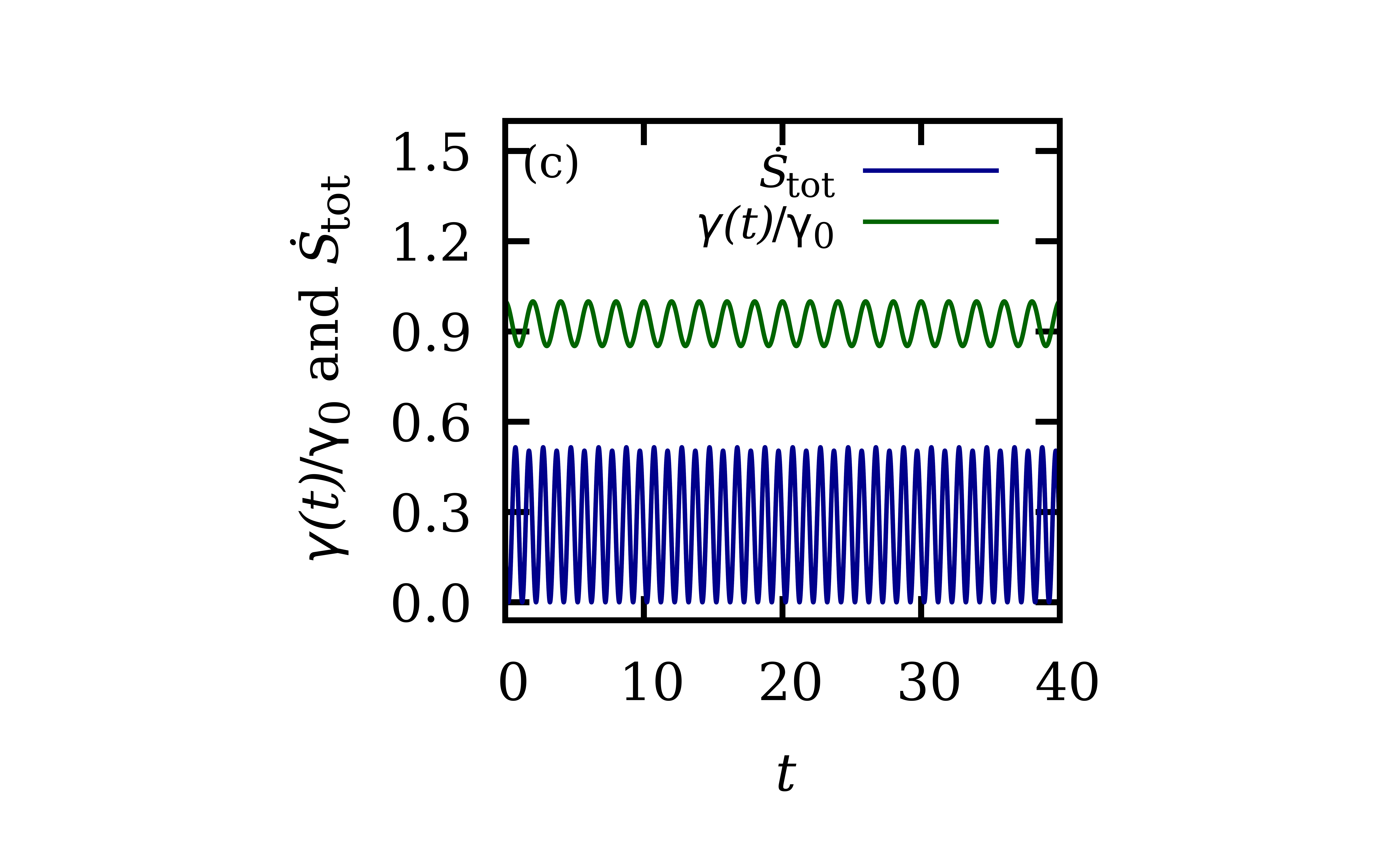} 
  \caption{(a-b) Illustration of the hysteresis loops of $J_{n}(t)$ and $A_{n}(t)$, respectively, as functions of the external drive $\gamma(t)$ at different driving frequencies $f$. (c) Demonstration of the time evolution of the total entropy production rate $\dot{S}_{tot} (t)$ together with the normalized external driving $\gamma(t)/\gamma_{0}$ for driving frequency $f = 0.5$, showing a characteristic double-peak structure in the entropy production rate, reflecting the complex nonequilibrium response of the system to the periodic modulation.}
        \label{fig:diagnose_loop}
\end{figure*}

\subsection{Instantaneous total entropy production rate}
To explore more about the nonequilibrium features through stochastic thermodynamics~\cite{Mondal2022}, we extend 
our analysis beyond the Shannon entropy evaluated in the previous subsections. In this 
subsection, we compute the total entropy production rate within a chemical master equation framework. It is calculated using transition probabilities and the probability flux as a function of time. 

According to the Schnakenberg formula \cite{Schnakenberg1976}, the time derivative of the system entropy, $\dot{S}_{\text{sys}}(t)$, can be decomposed into two contributions: 
\begin{equation}\label{total_EPR}
    \frac{dS_{\text{sys}}}{dt} = \Pi - \Phi.
\end{equation}
One is the total entropy production rate 
$\Pi$, which shows irreversibility in the system. The other quantity is $\Phi$, 
the rate of entropy change in the environment. 
Here, we assume the environment is an ideal reservoir~\cite{Tome2018}. 

From the chemical master equation, Eq.~(\ref{eq:master_eq}), the transition between nearest neighboring states $n \longleftrightarrow n + 1$ defines a probability current (flux) \cite{Shiraishi2023} which is 
\begin{equation}\label{eq: J_current}
\begin{aligned}
    J_{n} (t) = t^{+}_n(t)\,P_n(t) - t^{-}_{n+1}(t)\,P_{n+1}(t)\\
\end{aligned}
\end{equation}
Using the chemical master equation, the time evolution of the system entropy can be written in terms of probability currents as \cite{Schnakenberg1976, Esposito2010} 
\begin{equation}\label{eq: Sys_eq}
\begin{aligned}   
    \dot S_{\text{sys}}(t) = \sum_{n} J_{n} (t) \ln \frac{P_{n}(t)}{P_{n+1}(t)}\\
\end{aligned}
\end{equation}
While the system entropy reflects the internal probability distributions, the irreversibility at the trajectory level arises from the asymmetry between forward and backward transition rates. This asymmetry is quantified by the medium entropy production, defined through the logarithm of the ratio of forward to backward transition rates~\cite{Schnakenberg1976, Xiao2008}:
\begin{equation}\label{eq: Env_eq}
    \Phi =  \sum_{n} J_{n} (t) \ln \frac{t^+_{n}(t)}{t^-_{n+1}(t)}.
\end{equation}
The expression for the total entropy production rate is
\begin{equation}\label{eq: EPR_hys}
    \Pi =\sum_{n=0}^{n_{\max}-1}J_n(t) \ln\left(\frac{t_n^{+}(t)\,P_n(t)}{t_{n+1}^{-}(t)\,P_{n+1}(t)}\right).
\end{equation}
We use this expression to analyze the hysteretic response of the entropy production rate under periodic driving.

The total entropy production rate (EPR) is evaluated using three equivalent methods: (1) Direct expression for EPR, solving Eq.~(\ref{eq: EPR_hys}) that is based on direct flux and transition probabilities~\cite{Schnakenberg1976}; (2) EPR calculated by separate contributions from system and medium $\dot{S}_{\text{sys}}(t) = \dot{S}_{\text{tot}}(t) - \dot{S}_{\text{med}}(t)$, solving Eq.~(\ref{eq: Sys_eq} and \ref{eq: Env_eq}) \cite{{Das2012}}; (3) EPR is evaluated using the time derivative of the system entropy derived from Eq.~(\ref{eq:shannon_eq_CME}), where the derivative is calculated numerically using a central difference scheme, and combined with the medium entropy $\Phi$ flow term from Eq.~(\ref{eq: Env_eq}). All these approaches give similar results, which validate the numerical implementation. The data obtained using method (1) have been presented in Fig.~\ref{fig:instantaneous_EPR}(a) and (b). The total entropy production rate is evaluated from the product of the probability current and the corresponding thermodynamic affinity, summed over all transitions in the network. 

With these data, the dynamic hysteresis behavior of the entropy production rate is examined. 
The simulations are carried out for the same parameter values listed in Table~(\ref{tab:parameters}). Fig.~(\ref{fig:instantaneous_EPR})(a) demonstrates hysteresis loops observed for the total entropy production rate as a function of external periodic drive at different frequency values $f$. The hysteresis loops indicate that the total entropy production rate remains positive throughout. This is consistent with the second law of thermodynamics. The lag in the total entropy production rate arises from the internal structure: probability flux $J_{n}(t)$ and affinity $A_{n}(t)$. Both thermodynamic terms depend on the probability distribution $P_{n}(t)$ of the number of autocatalytic species $\ce{X}$, which evolves according to the chemical master equation. 

Fig.~\ref{fig:instantaneous_EPR}(a) illustrates that at low frequency value $f = 0.3$, the driving is slow compared to the system's relaxation time, and the probability distribution closely tracks the instantaneous non-equilibrium steady state, producing a narrow hysteresis loop in the total entropy production rate. At intermediate driving frequencies $f =0.7-1.5$, the time scale of external modulation becomes comparable to the internal time scale of the reaction system. Therefore, the system responds maximally to the external control in this frequency regime. Consequently, broad hysteresis loops are observed for the entropy production rate in this frequency domain. In the high frequency regime $f = 2.0-3.0$, the external driving varies so fast that the system's internal reaction dynamics are unable to adjust to these changing conditions. This results in the formation of small hysteresis loops.

\begin{figure*}[ht]

\centering
    
    \includegraphics[width=0.495\linewidth, height=6cm]{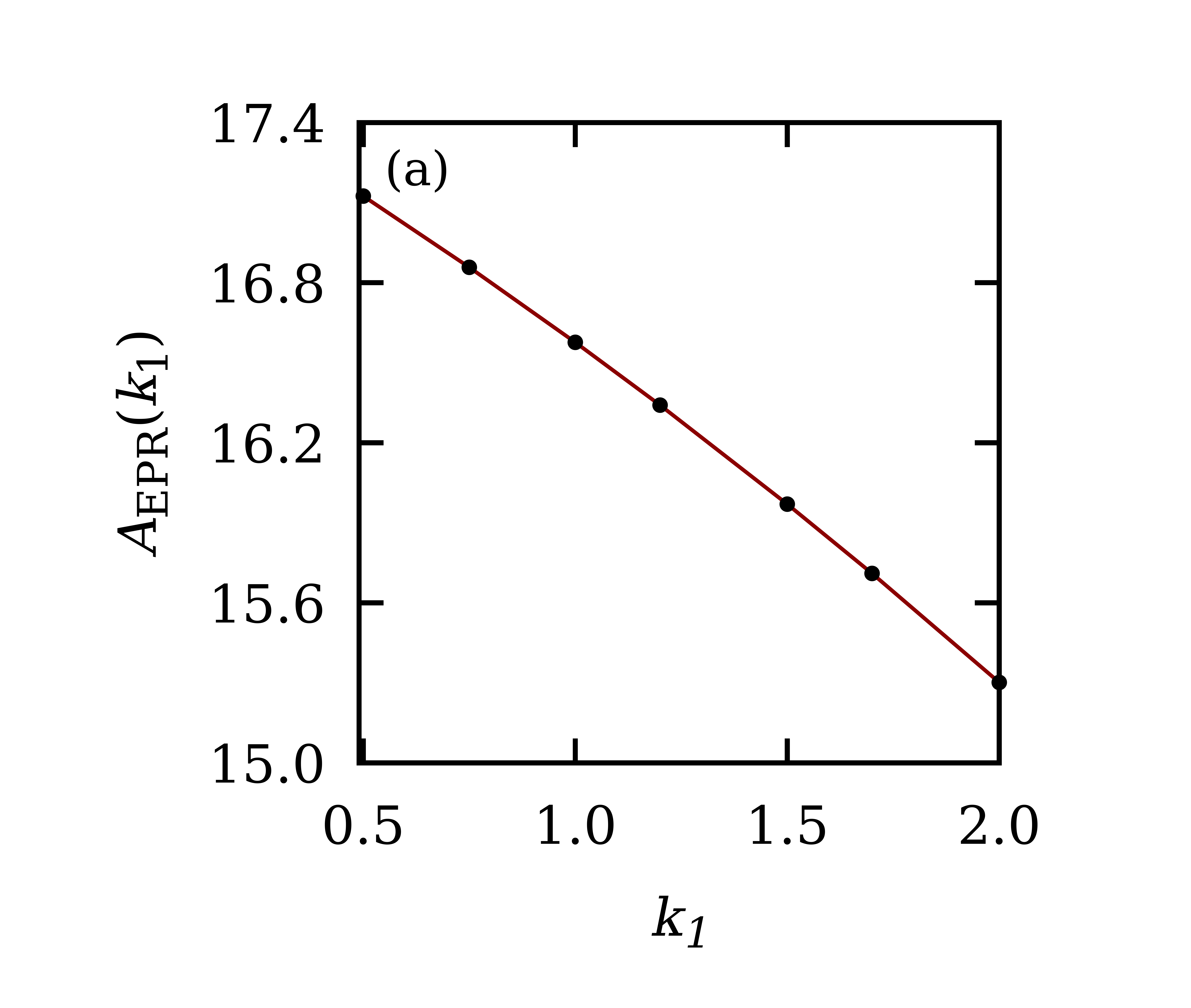}
    \includegraphics[width=0.495\linewidth, height=6cm]{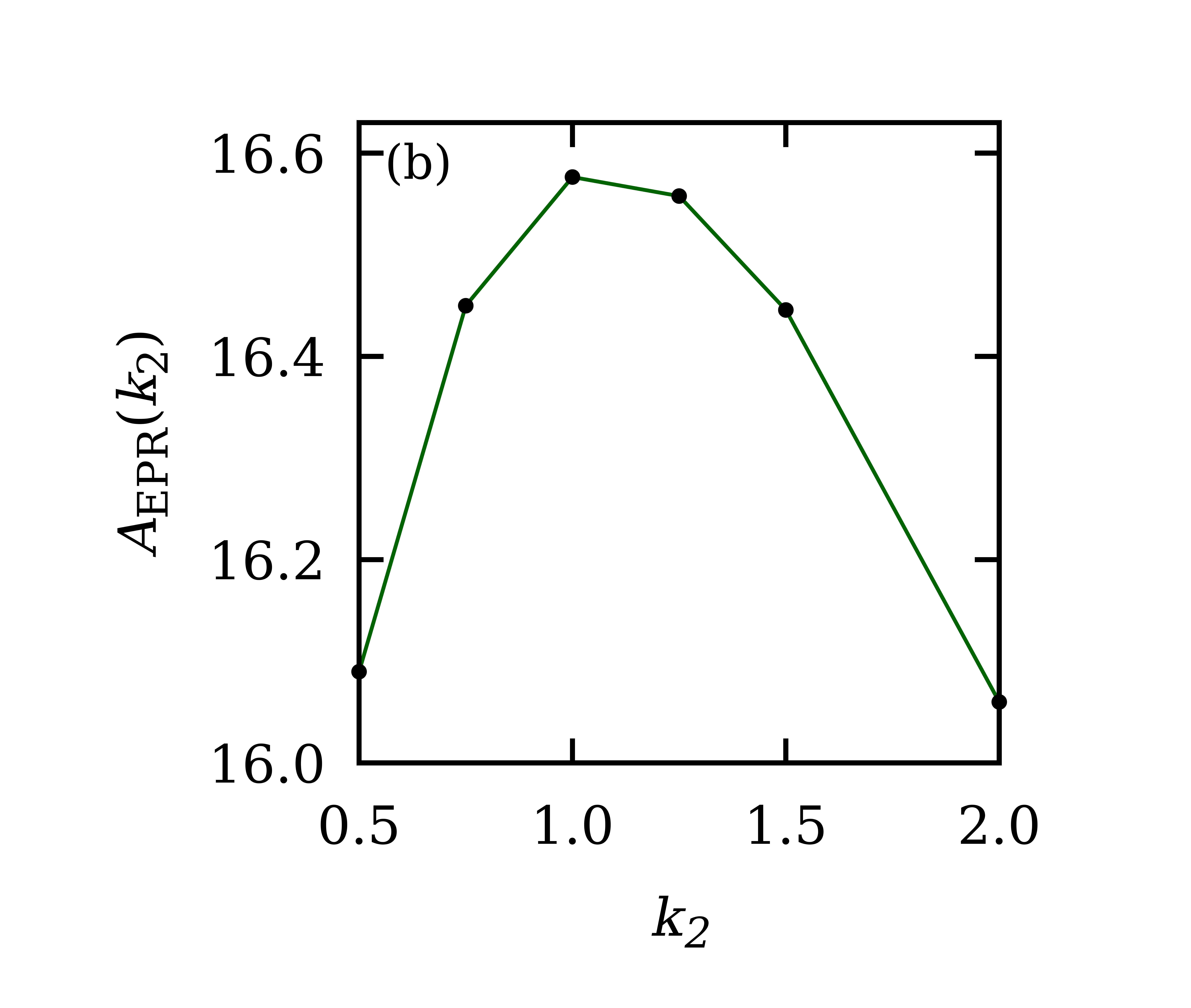}

\vspace{0.25cm}
  
    \centering

    \includegraphics[width=0.495\linewidth, height=6cm]{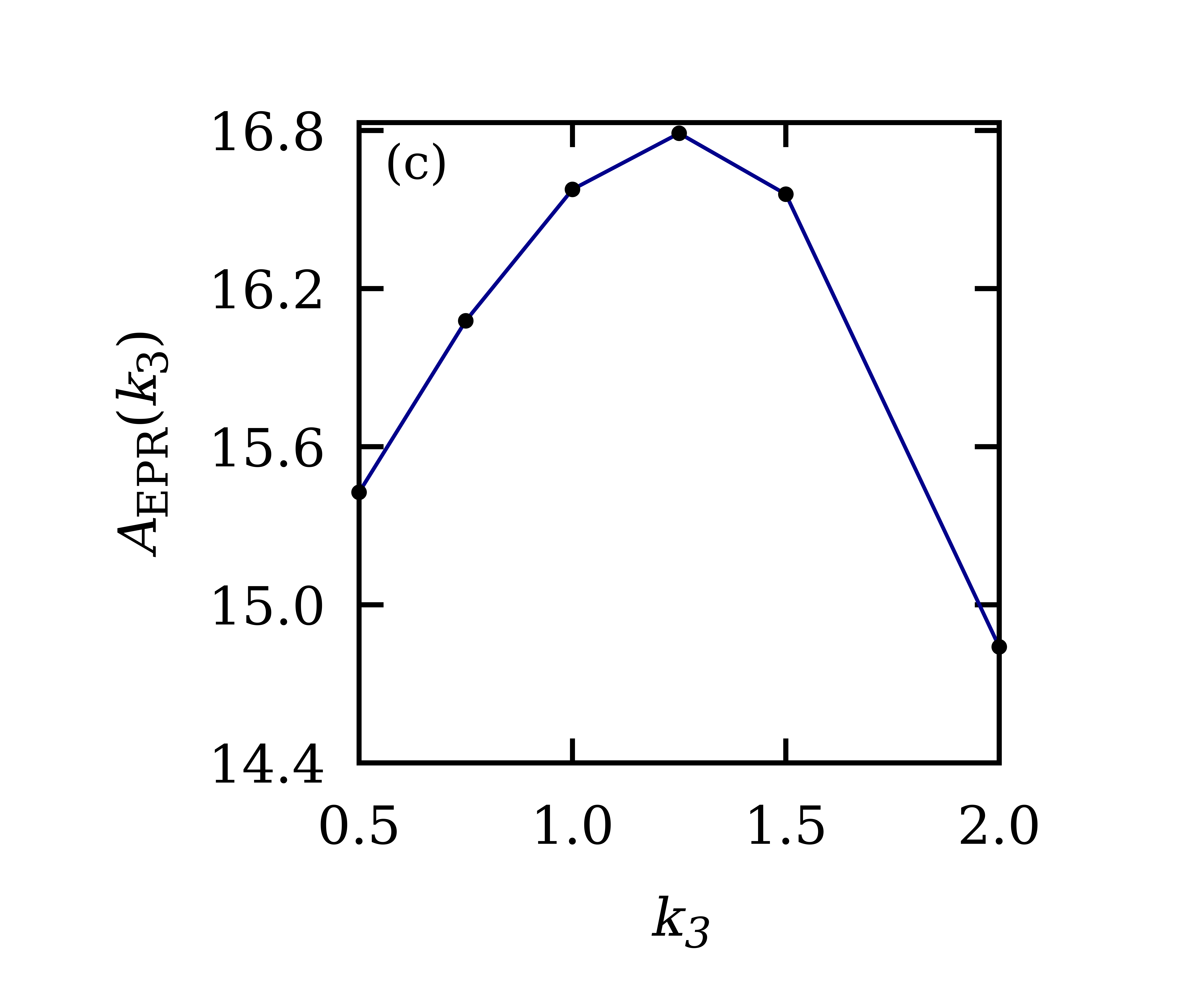}
    \includegraphics[width=0.495\linewidth, height=6cm]{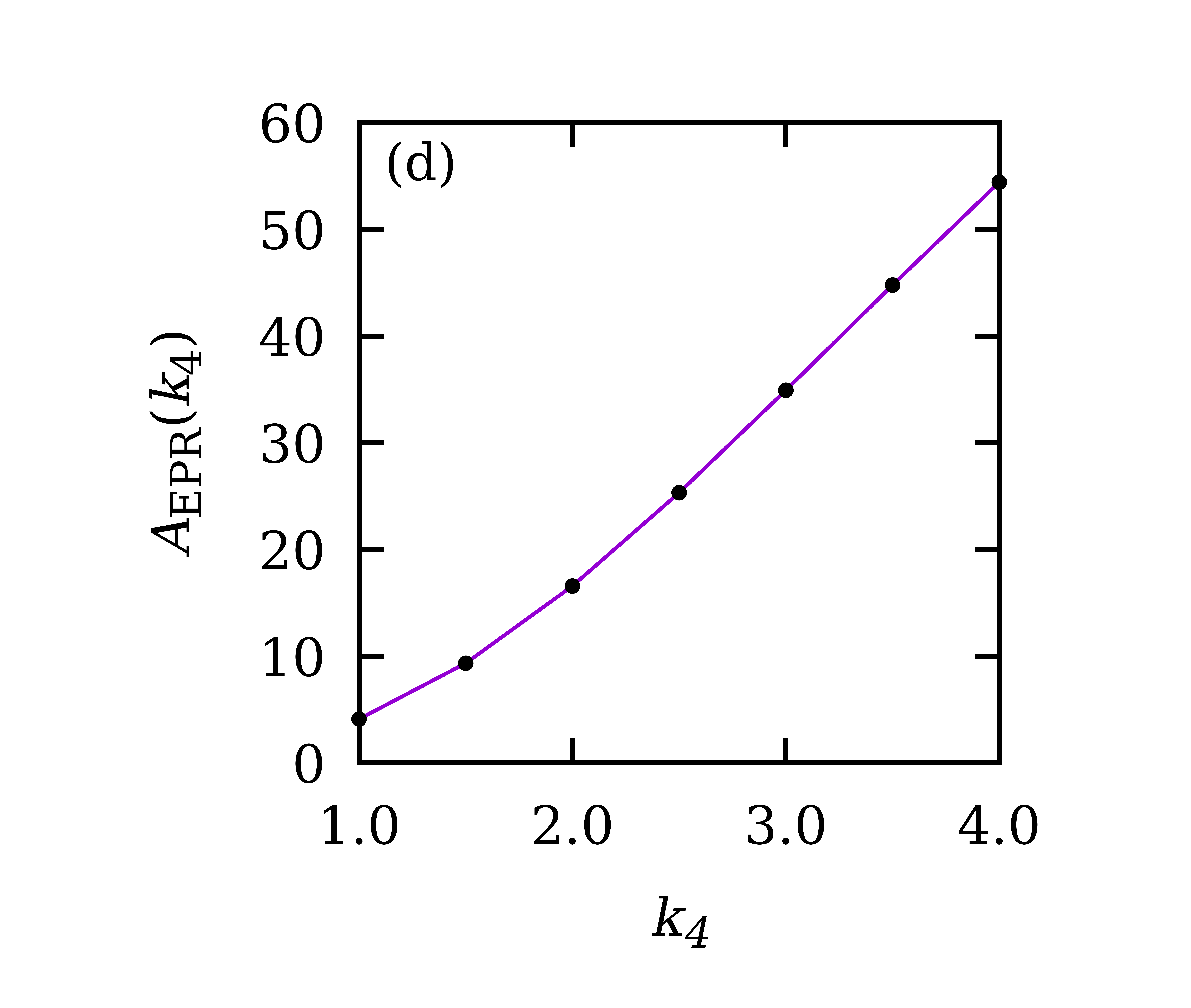}

        \caption{$A_{\textrm{EPR}}$ as the function of (a) $k_{1}$, (b) $k_{2}$, (c) $k_{3}$, and (d) $k_{4}$. For all cases, the concentrations have constant values with magnitudes of $A=0.3$, $B=2.7$, $C_{0}=1$, and $C_{1}=0.08$. The numeric value of the frequency of the external control,  $f=1$. The quantitative value of the volume of the system, $V = 100$. All quantities have dimensions as mentioned in Table I and Table II.}
        \label{fig:EPR_k}
\end{figure*}

To further quantify the observed hysteresis behavior across different frequency regimes, we analyze the corresponding entropy-production rate hysteresis loop area, denoted here by $A_{\mathrm{EPR}}(f)$. Fig.~\ref{fig:instantaneous_EPR}(b) shows the variation of $A_{\mathrm{EPR}}(f)$, the area enclosed by the hysteresis loop in entropy production rate, as a function of the driving frequency $f$ of the periodic control. 
At very low driving frequency, both flux and force remain in phase at each instant. The detailed balance condition is closely maintained. This results in a reaction dynamics close to equilibrium. This reduced extent of irreversibility and the occurrence of lower entropy production are reflected in small hysteresis loop areas. 
As the driving frequency increases to an intermediate regime, a proper lag between external modulation and thermodynamic response functions occurs. This imbalance in forward and backward rates enhances irreversible activity and dissipation. This gives rise to increased entropy production and maximum hysteresis loop area. At very high driving frequencies, the response of the system in terms of the thermodynamic flux is not much influenced by the external control due to its fast-changing nature. As a result, the degree of hysteresis behavior in the system decreases. Consequently, the irreversibility in the dynamics emerging because of that is reduced. This leads to lower dissipation, thereby reducing the hysteresis loop area. The maximum hysteresis loop area at intermediate frequency range $f=1.0$, for both Shannon entropy and entropy production rate, indicates that the system reaches an optimally driven state (optimal in terms of the hysteretic response resulting in hysteresis loss). The coincidence of these maxima shows consistency of the Maximum Entropy Production Principle (MEPP)~\cite{{Dobovivsek2018}}.  

The entropy production rate shows a similar variation of its hysteresis loop area with driving frequency as the Shannon entropy and the average concentration of the autocatalytic species. However, the shapes of the loops are distinct. It exhibits interesting butterfly shapes in response to the external periodic control. Generalized Preisach butterfly operator theory, in which butterfly loops arise due to competition between two opposite hysteretic contributions, gives a conceptual framework for the butterfly hysteresis loop for total entropy production rate \cite{VasquezBeltran2022}. Although our model is not described by a Preisach operator, this analogy suggests identifying the microscopic nonequilibrium contributions responsible. 

Given that the entropy production rate depends on the probability current and affinity, we concentrate on these two quantities to diagnose this butterfly shape in the hysteresis loop. We analyzed the time evolution and consequently, the hysteresis behavior of the total probability current flux $J_{n}(t)$ and the thermodynamic force or affinity $A_{n}(t) = \ln\left(\frac{t_n^{+}(t)\,P_n(t)}{t_{n+1}^{-}(t)\,P_{n+1}(t)}\right)$. Fig.~\ref{fig:diagnose_loop} (a) and (b) illustrate that the hysteresis loops observed for $J_{n}(t)$ and $A_{n}(t)$ are smooth, broad and elliptical in shape. In a given time-period, the periodic drive is maximally away from its mean value at two times. $|J_{n}(t)|$ and $|A_{n}(t)|$ reach their maxima twice. Consequently, the EPR shows two peaks in a single time period, illustrated in Fig.~\ref{fig:diagnose_loop}(c), which directly leads to the formation of the characteristic butterfly-shaped loop in the hysteresis representation. This double-peaked structure shows that dissipation is enhanced twice during each driving cycle, once in each half-cycle when the external modulation most strongly biases the system away from its instantaneous nonequilibrium state.


Finally, to analyze the effect of the chemical kinetic rate constants on the hysteresis loop area for the entropy production rate. 
We analyzed the variation of the hysteresis loop area for the entropy production rate $A_{\textrm{EPR}}$ as a function of the rate constants $k_{1}$, $k_{2}$, $k_{3}$, and $k_{4}$, individually. In analyzing the dependence of $A_{\textrm{EPR}}$ on one of the rate constants, we keep the magnitudes of all other rate constants fixed. Also, we maintained constant values of the concentrations, $A$, $B$, $C_{0}$, $C_{1}$ and the frequency of the external control $f$ for the measurements through numerical simulations. 

The rate constants have minimal effects on the quantity $A_{\textrm{EPR}}$, except $k_{4}$. The variation of $k_{4}$ significantly changes the magnitude of $A_{\textrm{EPR}}$ (Fig.~\ref{fig:EPR_k}).
This rate constant is directly associated with the periodic drive and appears in the transition probability expression (Eq.~(\ref{eq:birth_rate})). Since dynamic hysteresis results solely from the effect of the external periodic control, this particular analysis further supports the hysteresis here as dynamical. Moreover, an important interpretation regarding the equivalence of the description of the Langevin dynamics and the chemical master equations follows from this observation. In the transition probability expression (Eq.~(\ref{eq:birth_rate})), $k_{4}$ emerges as an amplitude term for the periodic drive. We detect that $A_{\textrm{EPR}}$ increases monotonically with $k_{4}$, analogous to the steady increase of $A_{\textrm{hys}}$ with respect to $\epsilon_{0}$, the amplitude factor in the Langevin framework.


\section{Conclusions}
\label{sec: conclusion}


We have shown that a driven autocatalytic chemical reaction network exhibits dynamic hysteresis. 
Schl\"ogl's model has the necessary underlying bistability through the autocatalytic species concentration. 
Driven by periodic pumping of the product species in the reaction medium, the autocatalytic species responds by oscillating between low and high concentrations under external periodic control. The hysteretic behavior manifests through the response lag, i.e., the concentration evolution of the autocatalytic species with respect to that of the product species. The results suggest that the effect of hysteresis tends to disappear in the limit of very slow external drive, supporting the dynamical nature of this hysteresis phenomenon. 

Analyzing this response across different relevant control parameters, including the frequency and amplitude of external driving, width of the potential well $w$, and volume of the system $V$, the dynamic hysteresis loop area exhibits turnover at an optimum driving frequency and well width. 
These observations imply that the system's maximum response to the external deterministic control can be tuned by adjusting the driving frequency, driving amplitude, well width, and system size. 

To connect the kinetic and thermodynamic descriptions, we analyzed dynamic hysteresis in Shannon entropy using both the Langevin equation and the chemical master equation. 
The dynamic hysteresis observed in the average concentration of the autocatalytic species $\langle x (t) \rangle$ and in the Shannon entropy $S(t)$ shows consistent behavior in both the hysteresis loops and the corresponding loop areas. 

To further elucidate the irreversibility of the system, we analyzed the dynamic hysteresis in the total entropy production rate. The agreement between the Shannon-entropy and entropy-production trends indicates that the delayed response is reflected both in state uncertainty and in thermodynamic dissipation. 
Therefore, within the present stochastic description, our integrated analysis shows that kinetic observables and thermodynamic observables provide mutually consistent signatures of dynamic hysteresis in this chemical system. 

Looking ahead, this phenomenon may have meaningful applications in the domain of the development of chemical computational networks. Autocatalytic reactions can be implemented to build fundamental chemical logic gates, the building blocks of chemical computers. Autocatalytic systems can provide bistability, which is required for logic operations, with the ``low'' and ``high'' concentration states of the autocatalytic species mapped to the ``off'' and ``on'' logical output states, respectively. 
The hysteresis here could ultimately be relevant to understand input-output correspondence, signal and threshold detection, and characteristic switching between output states for chemical logic gates. 

\section*{Acknowledgments}
S.Y. acknowledges IIT Mandi for a fellowship. M.D. thanks SERB (Project No. SRG/2022/000296), 
Department of Science and Technology, Government of India, 
and IIT Mandi (Seed Grant No. IITM/SG/MUD/91) for financial support. 
The High Performance Computing Cluster facility and Param Himalaya Supercomputing facility, 
managed by IIT Mandi, are also acknowledged. 
J.R.G was supported by the U.S. Department of Energy, Office of Science, Office of Basic Energy Sciences, Funding for Accelerated, Inclusive, Research (FAIR) under Award No. DE-SC-0024305.

\textbf{Author contributions:} M.D. and J.R.G. conceptualized and designed the research problem.
S.Y. executed the research work. S.Y., M.D. and J.R.G. analyzed the results. 
S.Y., M.D. and J.R.G. prepared and finalized the manuscript draft.
The authors express no conflicts of interest.


\begin{thebibliography}{99}

\bibitem{Remlein2024}
B.~Remlein and U.~Seifert,
Nonequilibrium fluctuations of chemical reaction networks at criticality: The Schl\"ogl model as paradigmatic case,
\textit{J. Chem. Phys.} \textbf{160}, 134103 (2024).

\bibitem{leonard1994stochastic}
D.~S. Leonard and L.~E. Reichl, Stochastic resonance in a chemical reaction,
\textit{Phys. Rev. E} \textbf{49}(2), 1734 (1994).

\bibitem{wong2022}
D. V. Kriukov, A. H. Koyuncu, and A. S. Y. Wong, History dependence in a chemical reaction network enables dynamic switching, 
\textit{Small} \textbf{18}, 2107523 (2022).





\bibitem{nicolaou2023prevalence}
Z. G. Nicolaou, S. B. Nicholson, A. E. Motter, and J. R. Green, Prevalence of multistability and nonstationarity in driven chemical networks,
\textit{J. Chem. Phys.} \textbf{158}, 224108 (2023).

\bibitem{delbruck1940}
M.~Delbr\"uck, Statistical fluctuations in autocatalytic reactions,
\textit{J. Chem. Phys.} \textbf{8}(1), 120--124 (1940).

\bibitem{nicolis1977}
G.~Nicolis and I.~Prigogine,
\textit{Self-Organization in Nonequilibrium Systems: From Dissipative Structures to Order through Fluctuations},
John Wiley \& Sons, New York (1977).


\bibitem{Vanag1999}
V.~K.~Vanag and G.~Nicolis,
Nonlinear chemical reactions in dispersed media:
The effect of slow mass exchange on the steady-state of the Schl\"ogl models,
\textit{J. Chem. Phys.} \textbf{110}, 4505 (1999).



\bibitem{mahato1994hysteresis}
M.~C. Mahato and S.~R. Shenoy, Hysteresis loss and stochastic resonance: A numerical study of a double-well potential,
\textit{Phys. Rev. E} \textbf{50}(4), 2503 (1994).

\bibitem{Das2012_MD}
M.~Das, D.~Mondal, and D.~S.~Ray,
Shape fluctuation-induced dynamic hysteresis,
\textit{J. Chem. Phys.} \textbf{136}, 114104 (2012).



\bibitem{casteels2016dynamic}
W.~Casteels, F.~Storme, A.~Le Boit\'e, and C.~Ciuti,
Power laws in the dynamic hysteresis of quantum nonlinear photonic resonators,
\textit{Phys. Rev. A} \textbf{93}, 033824 (2016).

\bibitem{chakrabarti1999dynamic}
B.~K.~Chakrabarti and M.~Acharyya,
Dynamic transitions and hysteresis,
\textit{Rev. Mod. Phys.} \textbf{71}(3), 847 (1999).




\bibitem{Dykman1994PRL}
M. I. Dykman, J. M. F. de Souza, M. L. E. Machado,
Hysteresis and noise-induced transitions in periodically driven systems,
\textit{Phys. Rev. Lett.} \textbf{72}, 1304 (1994).

\bibitem{Gammaitoni1998}
L.~Gammaitoni, P.~H\"anggi, P.~Jung, and F.~Marchesoni,
Stochastic resonance,
\textit{Rev. Mod. Phys.} \textbf{70}, 223 (1998).

\bibitem{Ghosh2025}
S.~Ghosh and M.~Das,
Theory of stochastic resonance with state-dependent diffusion,
\textit{Phys. Rev. E} \textbf{111}, 014125 (2025).

\bibitem{Paul2021}
S.~Paul, G.~Kotagiri, R.~Ganguly, H.~Courtois, C.~B.~Winkelmann, and A.~K.~Gupta,
Stochastic Resonance in Thermally Bistable Josephson Weak Links and Micro-SQUIDs,
\textit{Phys. Rev. Applied} \textbf{15}, 024009 (2021). 

\bibitem{ross1976} C. L. Creel and J. Ross, Multiple stationary states and hysteresis in a chemical reaction, 
\textit{J. Chem. Phys.} \textbf{65}, 3779 (1976).

\bibitem{Pal2025}
S.~Pal, M.~Panigrahy, R.~Adhikari, and A.~Dua,
Memory, hysteresis, and kinetic cooperativity in stochastic mnemonic networks,
\textit{J. Chem. Phys.} \textbf{162}, 125102 (2025).

\bibitem{guidi1997bistability}
G.~M. Guidi and A. Goldbeter, Bistability without hysteresis in chemical reaction systems,
\textit{J. Phys. Chem. A} \textbf{101}(49), 9367--9376 (1997).

\bibitem{Guidi1998}
G.~M.~Guidi and A.~Goldbeter,
Bistability without hysteresis in chemical reaction systems: The case of nonconnected branches of coexisting steady states,
\textit{J. Phys. Chem. A} \textbf{102}, 7813--7820 (1998).

\bibitem{Schiffmann1982}
Y.~Schiffmann,
Chemical triggering and hysteresis: The privileged position of two-dimensional chemical systems,
\textit{Physica A: Statistical Mechanics and its Applications} \textbf{114}, 74--83 (1982).

\bibitem{Ball2001}
R.~Ball and A.~D.~J.~Haymet,
Bistability and hysteresis in self-assembling micelle systems: phenomenology and deterministic dynamics,
\textit{Phys. Chem. Chem. Phys.} \textbf{3}, 4753--4761 (2001).

\bibitem{pajaro2019transient}
M.~P\'ajaro, I.~Otero-Muras, C.~V\'azquez, and A.~A.~Alonso,
Transient hysteresis and inherent stochasticity in gene regulatory networks,
\textit{Nat. Commun.} \textbf{10}, 4581 (2019).

\bibitem{Kim2012}
J. Kim, K. Kwon, and K. H. Cho,
The regulatory circuits for hysteretic switching in cellular signal transduction pathways,
\textit{FEBS J.} \textbf{279}, 878--887 (2012).

\bibitem{Aguda2003}
B. D. Aguda, Y. G. Kim, M. G. Piper-Hunter, A. Friedman, and C. B. Marsh,
Theoretical and experimental evidence for hysteresis in cell proliferation,
\textit{Cell Cycle} \textbf{2}(2), 167--171 (2003).

\bibitem{jiang2019single}
Y.~Jiang, X.~Li, B.~R.~Morrow, A.~Pothukuchy, J.~Gollihar, R.~Novak, C.~B.~Reilly, A.~D.~Ellington, and D.~R.~Walt, Single-molecule mechanistic study of enzyme hysteresis,
\textit{ACS Cent. Sci.} \textbf{5}(10), 1691--1698 (2019).

\bibitem{wang2023model}
T.~Wang, M.~Noori, W.~A.~Altabey, Z.~Wu, R.~Ghiasi, S.-C.~Kuok, A.~Silik, N.~S.~D.~Farhan, V.~Sarhosis, and E.~N.~Farsangi,
From model-driven to data-driven: A review of hysteresis modeling in structural and mechanical systems,
\textit{Mech. Syst. Signal Process.} \textbf{204}, 110785 (2023).

\bibitem{lynch2025hysteresis}
S.~T.~Lynch and S.~Lynch,
Hysteresis in neuron models with adapting feedback synapses,
\textit{Applied Math} \textbf{5}(2), 70 (2025).

\bibitem{Tierno2013}
P. Tierno, T. H. Johansen, and J. M. Sancho,
Unconventional dynamic hysteresis in a periodic assembly of paramagnetic colloids,
\textit{Phys. Rev. E} \textbf{87}, 062301 (2013).

\bibitem{das2012dynamical}
D. Das, M. Das, and D.~S. Ray, Dynamical hysteresis in a self-oscillating polymer gel,
\textit{J. Chem. Phys.} \textbf{137}(6) (2012).

\bibitem{Banerjee2015}
K.~Banerjee,
Dynamic memory of a single voltage-gated potassium ion channel: A stochastic nonequilibrium thermodynamic analysis,
\textit{J. Chem. Phys.} \textbf{142}, 185101 (2015).

\bibitem{sun2026universal}
Y.~Sun, X.~Li, Y.~Wang, J.~Zhou, H.~Bai, and Y.~Jin,
Universal scaling laws for dynamical-thermal hysteresis,
arXiv:2603.24007 (2026).

\bibitem{Rao1990}
M. Rao, H. R. Krishnamurthy, and R. Pandit,
Magnetic hysteresis in two-dimensional Ising systems driven by oscillating fields,
\textit{Phys. Rev. B} \textbf{42}, 856--884 (1990).

\bibitem{Jo2007}
J. Y. Jo, H. S. Han, J.-G. Yoon, T. K. Song, S. H. Kim, and T. W. Noh,
Domain switching kinetics in disordered ferroelectric thin films,
\textit{Phys. Rev. Lett.} \textbf{99}, 267602 (2007).

\bibitem{Tagantsev2001}
A. K. Tagantsev, I. Stolichnov, E. L. Colla, and N. Setter,
Polarization fatigue in ferroelectric films: Basic experimental findings, phenomenological scenarios, and microscopic features,
\textit{J. Appl. Phys.} \textbf{90}(3), 1387--1402 (2001).

\bibitem{Bisquert2023}
J. Bisquert,
Hysteresis in organic electrochemical transistors: Distinction of capacitive and inductive effects,
\textit{J. Phys. Chem. Lett.} \textbf{14}(49), 10951--10958 (2023).

\bibitem{Bisquert2024PRX}
J. Bisquert,
Inductive and capacitive hysteresis of current-voltage curves: Unified structural dynamics in solar energy devices, memristors, ionic transistors, and bioelectronics,
\textit{PRX Energy} \textbf{3}(1), 011001 (2024).

\bibitem{AlBender2004}
F. Al-Bender, W. Symens, J. Swevers, and H. Van Brussel,
Theoretical analysis of the dynamic behavior of hysteresis elements in mechanical systems,
\textit{Int. J. Non-Linear Mech.} \textbf{39}(10), 1721--1735 (2004).

\bibitem{Danilin2017}
A. N. Danilin,
Vibrations of mechanical systems with energy dissipation hysteresis,
\textit{Mech. Solids} \textbf{52}(3), 254--265 (2017).

\bibitem{Laurson2012}
L. Laurson and M. J. Alava,
Dynamic Hysteresis in Cyclic Deformation of Crystalline Solids,
\textit{Phys. Rev. Lett.} \textbf{109}, 155504 (2012).




\bibitem{yamaguchi2021thermal}
M.~Yamaguchi,
Thermal hysteresis involving reversible self-catalytic reactions,
\textit{Acc. Chem. Res.} \textbf{54}(11), 2603--2613 (2021).

\bibitem{dadi2016}
R.~K.~Dadi, D.~Luss, and V.~Balakotaiah,
Dynamic hysteresis in monolith reactors and hysteresis effects during co-oxidation of CO and C$_2$H$_6$,
\textit{Chem. Eng. J.} \textbf{297}, 325--340 (2016).

\bibitem{raj2015steady}
R.~Raj, M.~P.~Harold, and V.~Balakotaiah,
Steady-state and dynamic hysteresis effects during lean co-oxidation of CO and C$_3$H$_6$ over Pt/Al$_2$O$_3$ monolithic catalyst,
\textit{Chem. Eng. J.} \textbf{281}, 322--333 (2015).

\bibitem{Edison2021}
J.~R.~Edison, R.~L.~Siegelman, Z.~Preisler, J.~Kundu, J.~R.~Long, and S.~Whitelam,
Hysteresis curves reveal the microscopic origin of cooperative CO$_2$ adsorption in diamine-appended metal–organic frameworks,
\textit{J. Chem. Phys.} \textbf{154}, 214704 (2021).

\bibitem{sarkisov2000hysteresis}
L.~Sarkisov and P.~A.~Monson,
Hysteresis in Monte Carlo and molecular dynamics simulations of adsorption in porous materials,
\textit{Langmuir} \textbf{16}(25), 9857--9860 (2000).

\bibitem{monson2012}
P.~A.~Monson,
Understanding adsorption/desorption hysteresis for fluids in mesoporous materials using simple molecular models and classical density functional theory,
\textit{Microporous Mesoporous Mater.} \textbf{160}, 47--66 (2012).

\bibitem{contreras2016specific}
L.~Contreras, J.~Id\'igoras, A.~Todinova, M.~Salado, S.~Kazim, S.~Ahmad, and J.~A.~Anta,
Specific cation interactions as the cause of slow dynamics and hysteresis in dye and perovskite solar cells: a small-perturbation study,
\textit{Phys. Chem. Chem. Phys.} \textbf{18}(45), 31033--31042 (2016).

\bibitem{tress2015}
W.~Tress, N.~Marinova, O.~Inganäs, M.~K.~Nazeeruddin,
S.~M.~Zakeeruddin, and M.~Grätzel,
Understanding the rate-dependent current--voltage hysteresis, slow time component, and ageing in CH$_3$NH$_3$PbI$_3$ perovskite solar cells,
\textit{Energy Environ. Sci.} \textbf{8}, 995--1004 (2015).

\bibitem{snaith2014}
H.~J.~Snaith, A.~Abate, J.~M.~Ball, et al.,
Anomalous hysteresis in perovskite solar cells,
\textit{J. Phys. Chem. Lett.} \textbf{5}, 1511--1515 (2014).

\bibitem{richardson2016}
G.~Richardson, S.~E.~J.~O'Kane, D.~A.~Nieman, et al.,
Can slow-moving ions explain hysteresis in the current--voltage curves of perovskite solar cells?
\textit{Energy Environ. Sci.} \textbf{9}, 1476--1485 (2016).


\bibitem{Schlogl1972}
F. Schl\"ogl,
``Chemical Reaction Models for Non-Equilibrium Phase Transitions,''
\textit{Zeitschrift f\"ur Physik} \textbf{253}(2), 147--161 (1972).

\bibitem{prakash1997}
S.~Prakash and G.~Nicolis,
Dynamics of Schl\"ogl models on low-dimensional lattices,
\textit{J. Stat. Phys.} \textbf{86}(5--6), 1289--1310 (1997).

\bibitem{Wang2017}
T.~Wang and P.~Plecháč,
Parallel replica dynamics method for bistable stochastic reaction networks: Simulation and sensitivity analysis,
\textit{J. Chem. Phys.} \textbf{147}, 234110 (2017). 

\bibitem{verma2018stochastic}
M.~K. Verma, A. Kumar, and A. Pattanayak,
Stochastic bistable systems: Competing hysteresis and phase coexistence,
\textit{JETP} \textbf{127}(3), 549--557 (2018).

\bibitem{QuondamAntonio2022}
S.~Quondam~Antonio, F.~R.~Fulginei, G.~M.~Lozito, A.~Faba, A.~Salvini, V.~Bonaiuto, and F.~Sargeni,
Computing frequency-dependent hysteresis loops and dynamic energy losses in soft magnetic alloys via artificial neural networks,
\textit{Mathematics} \textbf{10}, 2346 (2022).

\bibitem{Kim2017}
C.~Kim, A.~Nonaka, J.~B.~Bell,
A.~L.~Garcia, and A.~Donev,
Stochastic simulation of reaction-diffusion systems: A fluctuating-hydrodynamics approach,
\textit{J. Chem. Phys.} \textbf{146}, 124110 (2017).

\bibitem{Ma2016}
L. Ma, X. Li, and C. Liu,
From generalized Langevin equations to Brownian dynamics and embedded Brownian dynamics,
\textit{J. Chem. Phys.} \textbf{145}, 114102 (2016).


\bibitem{Egbert2019}
M. Egbert, J.-S. Gagnon, and J. Pérez-Mercader,
From chemical soup to computing circuit: Transforming a contiguous chemical medium into a logic gate network by modulating its external conditions,
\textit{Journal of the Royal Society Interface} \textbf{16}(158), 20190190 (2019).

\bibitem{Arcadia2021}
C. E. Arcadia, A. Dombroski, K. Oakley, S. L. Chen, H. Tann, C. Rose, E. Kim, S. Reda, B. M. Rubenstein, and J. K. Rosenstein,
Leveraging autocatalytic reactions for chemical domain image classification,
\textit{Chemical Science} \textbf{12}, 5464--5472 (2021).

\bibitem{Peng2024}
Peng, Z., and Adam, Z. R.,
``Two mechanisms for the spontaneous emergence, execution, and reprogramming of chemical logic circuits,''
\textit{Chaos, Solitons \& Fractals}, \textbf{184}, 114955 (2024).

\bibitem{kriukov2024}
D. V. Kriukov, J. Huskens, and A. S. Y. Wong,
``Exploring the programmability of autocatalytic chemical reaction networks,''
\textit{Nature Communications} \textbf{15}, 8289 (2024).

\bibitem{Tshiprut2009}
Z.~Tshiprut and M.~Urbakh,
Exploring hysteresis and energy dissipation in single-molecule force spectroscopy,
\textit{J. Chem. Phys.} \textbf{130}, 084703 (2009).

\bibitem{Santamaria2014}
I.~Santamar\'ia-Holek and A.~P\'erez-Madrid,
Thermostatistical description of small systems in nonequilibrium conditions: Energy conversion and harvesting,
\textit{Phys. Rev. E} \textbf{89}, 012144 (2014).

\bibitem{Sasso2006}
C. P. Sasso, V. Basso, M. LoBue, and G. Bertotti,
Carnot cycle for magnetic materials: The role of hysteresis,
\textit{Physica B: Condensed Matter} \textbf{372}, 9--12 (2006).

\bibitem{Li2017ECM}
D. Li, H. Wang, Y. Qin, L. Han, X. Wei, and D. Qin,
Entropy production analysis of hysteresis characteristic of a pump-turbine model,
\textit{Energy Convers. Manage.} \textbf{149}, 175--191 (2017).

\bibitem{gillespie1977exact}
D.~T.~Gillespie, Exact stochastic simulation of coupled chemical reactions,
\textit{J. Phys. Chem.} \textbf{81}(25), 2340--2361 (1977).

\bibitem{vanKampen2007}
N. G. van Kampen,
\textit{Stochastic Processes in Physics and Chemistry},
3rd ed. (Elsevier, Amsterdam, 2007).


\bibitem{vellela2009stochastic}
M. Vellela and H. Qian, Stochastic dynamics and non-equilibrium thermodynamics of a bistable chemical system: the Schlögl model revisited,
\textit{J. R. Soc. Interface} \textbf{6}(39), 925--940 (2009).

\bibitem{Gillespie2000}
D. T. Gillespie,
``The Chemical Langevin Equation,''
\textit{J. Chem. Phys.} \textbf{113}(1), 297--306 (2000).

\bibitem{Doering2005}
C.~R.~Doering, K.~V.~Sargsyan, and P.~Smereka,
A numerical method for some stochastic differential equations with multiplicative noise,
\textit{Phys. Lett. A} \textbf{344}(2--4), 149--155 (2005).

\bibitem{Phillips2025}
E.~T.~Phillips, B.~Lindner, and H.~Kantz,
Stabilizing Role of Multiplicative Noise in Nonconfining Potentials,
\textit{Phys. Rev. Res.} \textbf{7}(2), 023146 (2025).

\bibitem{mcdermott2013dynamic}
D.~McDermott, J.~Amelang, C.~J.~Olson Reichhardt, and C.~Reichhardt, Dynamic regimes for driven colloidal particles on a periodic substrate at commensurate and incommensurate fillings,
\textit{Phys. Rev. E} \textbf{88}(6), 062301 (2013).

\bibitem{RodriguezGallo2021}
C. Rodríguez-Gallo, A. Ortiz-Ambriz, and P. Tierno,
``Degeneracy and hysteresis in a bidisperse colloidal ice,''
\textit{Phys. Rev. Research} \textbf{3}, 043023 (2021).

\bibitem{sides1998stochastic}
S.~W. Sides, P.~A. Rikvold, and M.~A. Novotny,
Stochastic hysteresis and resonance in a kinetic Ising system,
\textit{Phys. Rev. E} \textbf{57}(6), 6512 (1998).

\bibitem{rikvold2000dynamic}
P.~A. Rikvold, G.~Korniss, C.~J. White, M.~A. Novotny, and S.~W. Sides,
Dynamic phase transition and hysteresis in kinetic Ising models,
in \textit{Computer Simulation Studies in Condensed-Matter Physics XII} \textbf{63}, 105--119 (2000).



\bibitem{das2013chaos}
M. Das and D.~S. Ray, Chaos-induced dynamical hysteresis: Energetic and entropic barriers,
\textit{Phys. Rev. E} \textbf{87}(3), 032135 (2013).


\bibitem{chen2025coercivity}
M. Chen, X.-H. Zhao, and Y.-H. Ma, Coercivity Panorama of Dynamic Hysteresis,
\textit{arXiv} arXiv:2506.24035 (2025).


\bibitem{Pal2017}
K.~Pal, B.~Das, and G.~Gangopadhyay,
Nonequilibrium response of a voltage gated sodium ion channel and biophysical characterization of dynamic hysteresis,
\textit{Journal of Theoretical Biology} \textbf{415}, 113--124 (2017).

\bibitem{Zheng2026}
J. Zheng and Z. Lu,
“Nonlinear response relations and fluctuation-response inequalities for nonequilibrium stochastic systems,”
\textit{J. Chem. Phys.} \textbf{164}, 184114 (2026).





\bibitem{Lucarini2010}
V.~Lucarini, K.~Fraedrich, and F.~Lunkeit,
Thermodynamic analysis of snowball Earth hysteresis experiment:
Efficiency, entropy production and irreversibility,
\textit{Quarterly Journal of the Royal Meteorological Society}
\textbf{136}, 2--11 (2010).












\bibitem{seifert2005entropy}
U. Seifert,
Entropy production along a stochastic trajectory and an integral fluctuation theorem,
\textit{Phys. Rev. Lett.} \textbf{95}, 040602 (2005).

\bibitem{mcquarrie1967stochastic}
D.~A. McQuarrie,
``Stochastic approach to chemical kinetics,''
\textit{Journal of Applied Probability} \textbf{4}(3), 413--478 (1967).



\bibitem{Gardiner2009}
C. W. Gardiner,
Stochastic Methods: A Handbook for the Natural and Social Sciences,
4th ed. (Springer, Berlin, 2009).

\bibitem{shannon1948}
C.~E.~Shannon,
``A Mathematical Theory of Communication,''
\textit{Bell Syst. Tech. J.} \textbf{27}, 379--423 (1948).

\bibitem{Lesne2014}
A.~Lesne,
``Shannon entropy: A rigorous notion at the crossroads between probability,
information theory, dynamical systems and statistical physics,''
\textit{Math. Struct. Comput. Sci.} \textbf{24}, e240311 (2014).

\bibitem{seifert2012}
U.~Seifert,
``Stochastic thermodynamics, fluctuation theorems and molecular machines,''
\textit{Rep. Prog. Phys.}\textbf{75}, 126001 (2012).



\bibitem{Mondal2022}
S. Mondal, J. S. Greenberg, and J. R. Green,
Dynamic scaling of stochastic thermodynamic observables for chemical reactions at and away from equilibrium,
\textit{J. Chem. Phys.} \textbf{157}, 194105 (2022).




\bibitem{Schnakenberg1976}
J.~Schnakenberg,
Network theory of microscopic and macroscopic behavior of master equation systems,
\textit{Rev. Mod. Phys.} \textbf{48}, 571 (1976).

\bibitem{Tome2018}
T. Tomé and M. J. de Oliveira,
``Stochastic thermodynamics and entropy production of chemical reaction systems,''
\textit{J. Chem. Phys.} \textbf{148}, 224104 (2018).

\bibitem{Shiraishi2023}
N.~Shiraishi,
\textit{An Introduction to Stochastic Thermodynamics: From Basic to Advanced},
Fundamental Theories of Physics Vol.~212 (Springer, Singapore, 2023).

\bibitem{Esposito2010}
M.~Esposito and C.~Van den Broeck,
Three faces of the second law. I. Master equation formulation,
\textit{Phys. Rev. E} \textbf{82}, 011143 (2010).

\bibitem{Xiao2008}
T. J. Xiao, Z. Hou, and H. Xin,
Entropy production and fluctuation theorem along a stochastic limit cycle,
\textit{J. Chem. Phys.} \textbf{129}, 114506 (2008).

\bibitem{Das2012}
B.~Das, K.~Banerjee, and G.~Gangopadhyay,
Entropy hysteresis and nonequilibrium thermodynamic efficiency of ion conduction in a voltage-gated potassium ion channel,
\textit{Phys. Rev. E} \textbf{86}, 061915 (2012).

\bibitem{Dobovivsek2018}
A.~Dobovi{\v{s}}ek, R.~Markovi{\v{c}}, M.~Brumen, and A.~Fajmut,
The maximum entropy production and maximum Shannon information entropy in enzyme kinetics,
\textit{Physica A} \textbf{496}, 220--232 (2018).

\bibitem{VasquezBeltran2022}
M.~A.~Vasquez-Beltran, B.~Jayawardhana, and R.~Peletier,
On the Characterization of Butterfly and Multiloop Hysteresis Behavior,
\textit{IEEE Trans. Autom. Control }\textbf{67}, 3494--3506 (2022).

\bibitem{Qian1988}
S. Qian and D. O. Northwood,
Hysteresis in metal--hydrogen systems: A critical review of the experimental observations and theoretical models,
\textit{Int. Mater. Rev.} \textbf{33}(1), 25--52 (1988).

\bibitem{alsoubaihi2018}
R. M. Al Soubaihi, K. M. Saoud, and J. Dutta,
``Critical Review of Low-Temperature CO Oxidation and Hysteresis Phenomenon on Heterogeneous Catalysts,''
\textit{Catalysts} \textbf{8}(12), 660 (2018).

\bibitem{chen2017contact}
S.-Y. Chen, Y. Kaufman, A.~M. Schrader, D. Seo, D.~W. Lee,
S.~H. Page, P.~H. Koenig, S. Isaacs, Y. Gizaw, and J.~N. Israelachvili,
``Contact Angle and Adhesion Dynamics and Hysteresis on Molecularly Smooth Chemically Homogeneous Surfaces,''
\textit{Langmuir} \textbf{33}, 10041--10050 (2017).








\end{thebibliography}
\end{document}